\documentclass[useAMS,usenatbib,graphicx]{mn2e}

\usepackage{times}
\usepackage{graphicx}
\usepackage{pifont}        
\usepackage{color}

\title[Stellar feedback in AGN tori]{The effect of stellar feedback on the 
       formation and evolution of gas and dust tori in AGN}

\author[M. Schartmann et al.]
  {M.~Schartmann,$^{1,2,3,}$\thanks{E-mail: schartmann@mpe.mpg.de} 
   K.~Meisenheimer,$^{1}$ 
   H.~Klahr,$^{1}$
   M.~Camenzind,$^{4}$
   S.~Wolf$^{1,5}$ 
   \newauthor
   and Th.~Henning$^{1}$\\
$^{1}$Max-Planck-Institut f\"ur Astronomie, K\"onigstuhl 17, D-69117 Heidelberg, Germany\\
$^{2}$Max-Planck-Institut f\"ur extraterrestrische Physik, Giessenbachstra\ss e, D-85748 Garching, Germany\\
$^{3}$Universit\"ats-Sternwarte M\"unchen, Scheinerstra\ss e 1, D-81679 M\"unchen, Germany\\
$^{4}$ZAH, Landessternwarte Heidelberg, K\"onigstuhl 12, D-69117 Heidelberg, Germany \\
$^{5}$Christian-Albrechts-Universit\"at zu Kiel, Leibnizstra\ss e 15, D-24098 Kiel, Germany}
\begin{document}

\date{Accepted . Received ; in original form }

\pagerange{\pageref{firstpage}--\pageref{lastpage}} \pubyear{2008}

\maketitle
\begin{abstract}
Recently, the existence of geometrically thick dust structures in Active Galactic Nuclei (AGN)
          has been directly proven with the help 
          of interferometric methods in the mid-infrared. The observations are consistent with
          a two-component model made up of a geometrically thin and warm 
          central disk, surrounded by a colder, 
          fluffy torus component.
         Within the framework of an exploratory study, we investigate one possible 
          physical mechanism, which could produce such a structure, namely the 
          effect of stellar feedback from a young nuclear star cluster
          on the interstellar medium in centres of AGN.
         The model is realised by numerical simulations with the help of the hydrodynamics 
          code {\sc TRAMP}. We follow the evolution of the interstellar medium by taking discrete mass loss 
          and energy ejection due to stellar processes, as well as optically thin radiative 
          cooling into account. In a post-processing step, we calculate observable quantities like
          spectral energy distributions and surface brightness distributions with the help of the 
          radiative transfer code {\sc MC3D}.
         The interplay between injection of mass, supernova explosions and radiative cooling
          leads to a two-component structure made up of a cold geometrically thin, but optically 
          thick and 
          very turbulent disk residing in the vicinity of the angular momentum barrier, 
          surrounded by a filamentary structure. The latter consists of 
          cold long radial filaments flowing towards the disk and a hot tenuous medium in 
          between, which shows both inwards and outwards directed motions.
          With the help of this modelling, we are able to reproduce the range of observed 
	  neutral hydrogen column 
          densities of a sample of Seyfert galaxies as well as the relation between 
          them and the strength of the 
          silicate 10\,$\umu$m spectral feature. 
          Despite being quite crude, our mean Seyfert galaxy model is even able 
          to describe the SEDs of two intermediate type Seyfert galaxies observed with the
          {\it Spitzer} Space Telescope.
\end{abstract}

\label{firstpage}

\section{Introduction}
\label{sec:intro}

Within the so-called {\it Unified Scheme of Active Galactic Nuclei} \citep{Antonucci_93,Urry_95}, a
torus-like geometrically thick gas and dust component is 
coined to explain two observed classes of Active Galactic Nuclei (AGN) by one intrinsically unique AGN model.
This torus is illuminated by the accretion disk, surrounding the central black 
hole. As the dust opacity peaks in the UV/optical wavelength range, 
where also the accretion disk emits maximally, most of its light is absorbed and reemitted 
in the form of a pronounced peak in the mid-infrared \citep{Sanders_89}, which 
is frequently observed in AGN. This gives rise to the separation into type~2 sources, 
where the torus is viewed edge-on and type~1 sources, where a face-on view  
allows the visibility of the accretion disk, which shows up in form of the so-called 
{\it Big Blue Bump} in the spectral energy distribution (SED) of AGN. 
This geometrical unification idea was first proposed by \citet{Antonucci_85}, 
after broad emission lines
-- arising from fast moving gas close to the central black hole (the so-called 
{\it Broad Line Region}) -- have been
detected in polarised light in the Seyfert~2 galaxy NGC~1068, which was explained by
scattering by dust and electrons above the torus opening. 
In direct light, type~2 sources only exhibit narrow emission lines, originating from
slower gas motions further out, where they are less affected by the gravitational potential. 
This is the so-called {\it Narrow Line Region}, located beyond the opening 
of the torus funnel. 
The first direct evidence for a dusty torus came from interferometric observations
with the help of the MID-infrared Interferometer (MIDI),
which revealed geometrically
thick dust distributions on parsec scale in the two Seyfert galaxies NGC\,1068
\citep{Jaffe_04,Poncelet_06,Raban_08}
and the Circinus galaxy \citep{Tristram_07}.

Up to now, no conclusive, physical model for the distribution of gas and dust within 
these tori exists. 
This is due to the fact that very complicated and yet not fully understood 
astrophysical processes are 
thought to happen in the centres of AGN. 
Additionally, their large distances do not allow direct imaging observations.
The major problem for the persistence of geometrically thick gas and dust
distributions is to obtain stability of the vertical scale height against gravity. 
So far, several models have been put forward: 
\citet{Krolik_88} proposed that
the torus is made up of clumps which possess supersonic random velocities,
maintained by transferring orbital shear energy with the help of sufficiently elastic
collisions between the clumps (see also \citealp{Beckert_04}). 
To reach this elasticity, high magnetic field
strengths are necessary. 
As these tori are made up of a multiphase mixture of gas and dust with cold, 
warm and hot components, a clumpy structure also helps to prevent the dust 
from being destroyed by hot surrounding gas. Further evidence for clumpiness 
or a filamentary structure of the cold component -- in this case mainly for the 
distribution of neutral gas -- comes from X-ray measurements 
of the absorbing column density distribution \citep{Risaliti_02} and their combination
with measurements of the strength of a spectral feature of a dust ingredient (see also discussion
in Sect.~\ref{sec:feat_hi}).
Very recently, \citet{Tristram_07} found hints for clumpiness in the dust torus of the Circinus
galaxy by means of interferometric observations in the mid-infrared. 
\citet{Wada_02} performed hydrodynamical simulations of the interstellar medium in centres of 
active galaxies under the assumption of starburst conditions. A very high 
supernova type~II rate in a narrow sheet around the midplane (following in situ 
star formation in the densest region) is able to puff up 
an initially rotationally supported thin disk.   
In an alternative scenario, no torus is needed, but 
the necessary obscuration is given by dusty clouds, which are embedded 
into a hydromagnetic disk wind 
(\citealp{Koenigl_94} and discussion in \citealp{Elitzur_06}).  
The most recent approach comes from \citet{Krolik_07}, following an idea
of \citet{Pier_92}. His idealised analytical calculations show that 
the scale-height of AGN tori 
can be stabilised against gravity with the
help of infrared radiation pressure.   

A second approach towards modelling gas and dust structures of AGN is via 
radiative transfer calculations of specific dust distributions, which 
are motivated by simplified astrophysical scenarios. 
The most up-to-date simulations
\citep{Nenkova_02,Hoenig_06,Schartmann_08,Nenkova_08a,Nenkova_08b} feature clumpy dust distributions.
From these calculations, we get some rough idea of the geometrical distribution of dust within 
AGN from comparison with high resolution spectral (e.g.\/ NACO or {\it Spitzer}) 
as well as interferometric observations with MIDI (the MID-infrared Interferometer).

The goal of this work is to explore the effects of stellar feedback from a nuclear stellar 
cluster on the formation and evolution of AGN tori in terms of an exploratory study. 
In this scenario, the torus is mainly 
built up by gas loss due to stellar evolution. Energetic supernova explosions provide an 
effective structuring mechanism. 
To this purpose, we deploy three-dimensional hydrodynamic simulations. 
Eventually, this model should account for the
obscuration as well as the feeding of the central source.
In a second step, we convert the gas density distribution into a dust distribution,
which we then use for continuum radiative transfer calculations, 
which can later be compared to observational 
results.
In this work, we focus on low-luminosity AGN (Seyfert galaxies), as they are
more abundant in the local universe and, therefore,
their nuclear regions are accessible via current generation interferometric
instruments.  

After having discussed the main physical assumptions of our work in 
Sect.~\ref{sec:model}, we will briefly present the numerical method 
we use (Sect.~\ref{sec:numerics}). In Sect.~\ref{sec:results}, we describe the results of this 
exploratory study in terms of the evolution of our 
standard torus model. Furthermore, we characterise its final state and explain the results from a 
study of varying several model parameters. After a critical discussion (Sect.~\ref{sec:discussion}), 
we compare our torus simulations to recent observations of Seyfert galaxies
(Sect.~\ref{sec:comp_obs}) in terms of the spectral energy distributions of dust 
re-emission, as well as gas
extinction column densities and summarise our work (Sect.~\ref{sec:summary}).

\section{Physical assumptions of our model}
\label{sec:model}

The main assumption of our model is that a young star cluster exists in
those nuclei of AGN, which possess a torus. Evidence for this was 
given recently by observations of a number of 
Seyfert galaxies observed with adaptive optics techniques \citep{Davies_07} and 
for the special case of NGC\,1068 with the help of 
HST/NICMOS images (\citealp{Gallimore_03} and references therein), one of the closest
-- and therefore best studied -- Seyfert~2 galaxies.  
We model this nuclear stellar cluster in form of a Plummer profile
\citep{Plummer_11} of the gravitational potential 
and assume that it was built up during a short-duration
starburst\footnote{We assume a starburst duration of 40\,Myrs in the following.}, 
in concordance with the findings of \citet{Davies_07}, which forms 
a so-called 
{\it Coeval Stellar Population (CSP)}. 
The first phase of evolution of the stellar population 
is expected to be very violent, such that
supernova type~II explosions might evacuate most of the nuclear region from
gas and dust. During this phase, we do not expect that a stable gas and dust distribution
(torus) exists.  
According to our population synthesis
modelling with the starburst\,99-code \citep{Leitherer_99,Vazquez_05}, this violent
phase lasts for approximately 40 million years. Within about the same time period,
double systems of lower mass stars have been able to form. According to
models which include binary systems \citep{DeDonder_03}, the rate of supernova type~Ia
explosions -- arising from so-called double degenerate systems -- begins to rise
steeply and starts to dominate the input of energy. 
At the same time, mass injection should be dominated by the emission of planetary nebulae from stars
in the intermediate mass range (from 1.5 to 8$\,M_{\sun}$, \citealp{Kwok_05}). 

\subsection{Mass input}
\label{sec:mass_input}

Stars of all masses lose material during their
lifetime in form of winds of different strengths, mainly happening within
short periods (e.g.\/ at the tip of the Red Giant Branch). 
For the case of high-mass
stars, also supernova type\,II explosions have to be taken into account. 
The time dependent mass loss rate of a CSP can in principle be calculated with
the following integration over the initial stellar mass for a given 
(time dependent) mass loss rate of individual stars $\dot{m}(t,m_i)$ and 
initial mass function (IMF) $\psi(m_i)$:

\begin{eqnarray}
\label{equ:massloss}
  \dot{M}(t) = \int_{m_{min}}^{m_{max}}\, \dot{m}(t,m_i) \, \psi(m_i) \, dm_i.  
\end{eqnarray}

As both functions are not very well known, a number of
simplifying assumptions have to be applied. We use the model 
of \citet{Jungwiert_01}, which summarises the various phases of 
mass loss in one
delta peak for the intermediate mass stars (IMS) and the high mass stars (HMS)
and two delta peaks for low mass stars (LMS).
For the IMS, the delta peak represents the AGB ({\it Asymptotic Giant
Branch}) wind peak and for the HMS
the SNe explosions plus winds. For the LMS, it represents the wind loss at the
tips of the RGB ({\it Red Giant Branch}) and AGB.
The IMF is modelled according to \citet{Scalo_98} as a broken power law. 
Using the stellar initial-final mass relation
and the stellar lifetime-mass relation
from the Padua stellar evolutionary models
\citep{Bressan_93,Marigo_96}, an 
approximate, normalised mass loss rate function can be derived:
\begin{eqnarray}
  \dot{M}_{\mathrm{n}}(t) = \frac{5.55\cdot 10^{-2}}{t+5.04\cdot 10^6\,\mathrm{yr}},
\end{eqnarray}
where the subscript n denotes that it is normalised to the initial mass of the CSP.
This leads to a mass loss rate of $1.2\cdot
10^{-9}\,M_{\sun}\,yr^{-1}\,M_{\sun}^{-1}$ (normalised to a total stellar mass 
of $1\,M_{\sun}$) at the beginning of our simulations
(after 40\,Myr). 
Recent observations
tend to find rather top heavy IMFs in the centres of galaxies. 
This means that the number of
massive stars is increased compared to low-mass stars. It has been
shown for the centre of the Milky Way
\citep{Stolte_02,Stolte_05,Nayakshin_06a,Paumard_06}, the Andromeda galaxy
\citep{Bender_05} as well as more distant galaxies
\citep{Scalo_90,Elmegreen_05}.
Therefore, we use a five times larger value of $\dot{M}_{\mathrm{n}} = 6.0\cdot
10^{-9}\,M_{\sun}\,yr^{-1}\,M_{\sun}^{-1}$ in our standard model, which is kept  
constant over the whole evolution of the simulation. For simplification reasons, 
we assume that all of the mass is injected in form of planetary nebulae. 

Numerically, mass injection means a source term in the continuity
equation as well as the momentum equation, when taking the ejection velocity of the
gas into account. 
According to the mass injection rate, the number of planetary nebulae (PNe) to 
inject\footnote{As the main mass
input at the time of our simulations is given by the ejection of PNe,
we further refer to this mass input mechanism as the planetary
nebulae injection, although all mass loss effects mentioned above are
included.} into the whole model space is calculated. 
Residual mass is injected into the model space within the subsequent time step.
The position of the
ejecting star is chosen randomly following the distribution of stellar mass.  
The ejection velocity of the PN is chosen
such that it matches the velocity dispersion (assumed to be independent of position) 
plus the rotational velocity 
of the stars in the central star
cluster. As we cannot follow the evolution of single stars explicitly, 
all of the ejected planetary nebulae have the same gas mass $m_{\mathrm{PN}}$,
which corresponds to the mean mass by weighting the initial-final mass relation (given
by \citealp{Weidemann_00}) with the Initial Mass Function (IMF), which in this mass regime
is well approximated by a Salpeter-like function \citep{Salpeter_55}. 
This leads to an average mass of $2.2\,M_{\odot}$ per
planetary nebula, which we 
distribute homogeneously over $3^3$ cells surrounding the randomly chosen
position. 

A detailed modelling of the temperature structure of a planetary nebula is
beyond the reach of our current resolution, as we are not able to resolve 
the small-scale physical processes. As it will finally
thermally merge with the surrounding gas on short time-scales, we
choose the temperature to be the same as the gas in the cells where the PNe are
injected.

\subsection{Energy input}
\label{sec:en_input}

Our simulations start after roughly 40\,Myr, when the violent phase of
supernova~II explosions (e.g.~a $40\,M_{\sun}$
mass star lives only about 5\,Myr) has ended. At roughly the same time, the rate
of SN~Ia explosions begins to dominate the energy input into the ISM\footnote{This is only the
case, if the so-called {\it double-degenerate} scenario of two merging 
white dwarfs is realised in nature.} -- a result found in population synthesis 
modelling including binary systems \citep{DeDonder_03}.
According to today's theoretical understanding, SN\,Ia are thought
to be thermonuclear disruptions of accreting white dwarfs, which reach
the {\sc Chandrasekhar} mass ($\approx$$1.4\,M_{\sun}$), being the
maximum mass which can be supported by electron
degeneracy pressure. 
The determination of SN\,Ia rates for the considered nuclear stellar system 
in our simulations  
is complicated. Observations have
to rely on low number statistics. Additionally, only the stellar content of the
whole galaxy can be taken into account. 
A recent publication by \citet{Sullivan_06} 
describes the SN\,Ia rate for young and old stellar populations
and finds that it can be well represented by the sum of $5.3 \pm 1.1 \cdot
10^{-14}\,\mathrm{SNe}\,\mathrm{yr}^{-1} M_{\sun}^{-1}$ (old population of stars) and $3.9 \pm 0.7 \cdot 10^{-4} \, \mathrm{SNe}\,
\mathrm{yr}^{-1} (M_{\sun}\,yr^{-1})^{-1}$ (young
population of stars). 
The rate for the old stellar population is normalised to 1\,$M_{\sun}$ of
stars and the latter rate to a star
formation rate of 1\,$M_{\sun}\,\mathrm{yr}^{-1}$.

For an assumed duration of the starburst of $40\,Myr$, this
results in a normalised SN\,Ia rate of 
roughly $10^{-11}\,\mathrm{SNe}\,M_{\sun}^{-1}\,\mathrm{yr}^{-1}$ for our modelling. 
Due to the large uncertainties in theoretical derivation of supernova rates 
and as it is doubtful whether the observational
derivations are appropriate for our simulations, we consider the SN\,Ia rate a
more or less free parameter within our models. 
In an alternative starburst scenario for nuclei of Seyfert galaxies, \citet{Wada_02} use
a much larger rate of roughly 100 times our SN-rate, namely 1\,SN detonation per
year, restricted to a narrow strap around the midplane of their model space. 
This is motivated by violent star
formation within a dense gas disk in the galactic midplane, accounting for
SN\,II explosions.  
After a SN-rate parameter study, a value of   
$10^{-10}\,\mathrm{SNe}\,M_{\sun}^{-1}\,yr^{-1}$ was chosen for our standard model. 
It is assumed to be constant in our simulations, as our current period of time 
evolution (1.2\,Myrs)
is short compared to the expected evolutionary time scales of AGN tori, of the order of
100\,Myrs.  
From the supernova rate for the whole model space, the number of supernovae exploding in each
timestep can be determined. When
a supernova is launched, the position is determined randomly, but according to the underlying stellar
distribution. The energy is
injected into a single cell in the form of thermal energy. 
Additionally, a mass
of $1.4\,M_{\sun}$ is added to this cell, also ensuring numerical
stability. We use an efficiency
of $\eta_{\mathrm{SN}}=0.1$,  
reducing the input of thermal energy to $10^{50}\,$erg. This factor
accounts for energy losses in the early phases of evolution 
of the supernovae, which we cannot model at the current resolution.
The problem of radiating away most of the introduced thermal energy due to
vigorous cooling in very dense regions before a supernova shell can form
is mitigated by introducing a 
cooling delay of several timesteps. Direct modelling of an expanding shell 
as initial condition for a supernova explosion is not feasible at the
resolution of our current simulations.
These shock fronts efficiently dissipate their kinetic energy
when interacting with dense blobs or filaments or dense shock fronts of 
neighbouring supernova explosions, transferring their kinetic energy back to 
thermal energy. 
Similar procedures of energy input have been applied to many different 
astrophysical scenarios. 
For example \citet{Wada_02} also introduce the complete amount of supernova energy into
a single cell in form of thermal energy within a starbursting gaseous disk with even higher ambient densities compared 
to our simulations. 
Purely thermal energy is also introduced by \citet{Joung_06} for the case of the interstellar medium in our own 
galaxy, where the mean gas density is much lower compared to our computations. 
Given the aims of this exploratory study, we therefore think that our energy injection mechanism is an appropriate 
treatment of this feedback process.

\subsection{Optically thin cooling}
\label{sec:cooling}

\begin{figure}
 \centering
  \resizebox{1.0\hsize}{!}{\includegraphics{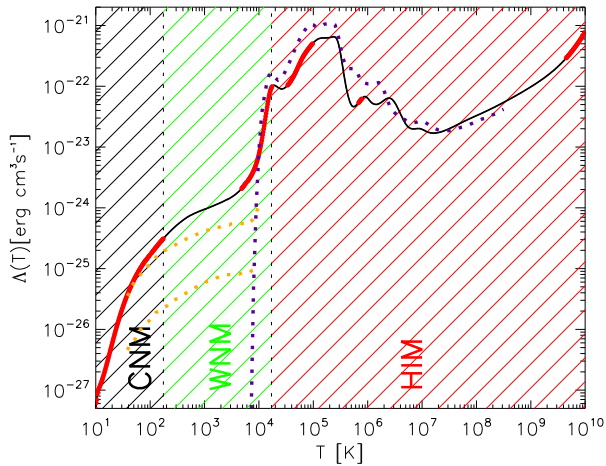}}
  \caption[Cooling curve]{Cooling curve \citep{Plewa_95}. The hatched regions denote three temperature
    regimes: cold neutral medium (CNM), warm neutral medium (WNM) and hot
    ionised medium (HIM). The thick red line segments mark the thermally stable parts of
    the cooling curve. Dotted lines refer to cooling curves used by other
    groups, see text.} 
  \label{fig:cool_curve} 
\end{figure}

A mixture of neutral and ionised gas mainly cools via the transformation of kinetic
energy to radiation by collisional processes. This also means that the
efficiency of cooling is a sensitive function of the constituents of the
gas. The radiation can only escape, if the medium is optically thin for these
wavelengths, otherwise cooling is inefficient.   

Fig.~\ref{fig:cool_curve} shows the effective radiative cooling curve applied in our
simulations \citep{Plewa_95}, which was calculated with the help of the 
{\it Cloudy} code \citep{Ferland_93}. 
For the case of our cooling curve, a purely
collisionally ionised optically thin plasma with solar abundances was used.   

Following \citet{Joung_06}, such a gas is thermally stable at temperatures, 
where the slope in logarithmic
scale is steeper or equal to one. For the case of our cooling
function, we find five distinct stable regions: $10-175\,$K, $4\,800-17\,400\,$K,
$33\,800-97\,200\,$K, $6.82\cdot 10^{5}-7.72\cdot 10^{5}\,$K and 
$4.47\cdot 10^9-1.00\cdot 10^{10}\,$K. 
They are marked with the thick red line segments in Fig.~\ref{fig:cool_curve}. 
Overplotted as yellow dotted lines are the often used cooling functions from
\citet{Dalgarno_72} -- where only the part below $10^4\,$K is shown -- 
with ionisation fractions of 0.1 for the
upper curve and 0.01 for the lower one. The violet dotted line results from
equilibrium ionisation cooling, calculated by \citet{Sutherland_93} for the
case of solar metallicity. 
As can be seen from this, the cooling curves are in relatively good agreement in the high
temperature regime (above the ionisation temperature of hydrogen at around $10^4\,$K).
At temperatures lower than $10^4\,$K, the values of the assumed cooling curves
are most uncertain, as in this temperature regime, the approximation of
thermal equilibrium is usually not valid. Therefore, large deviations exist
between the different calculational methods. It is also a matter of debate,
which ionisation fraction to assume for the low temperature
part. We decided to use 0.1 (black curve and upper yellow 
curve in Fig.~\ref{fig:cool_curve}) in concordance with
\citet{Avillez_04}, whereas \citet{Joung_06} assume an ionisation fraction of
0.01 (lower yellow curve). For temperatures lower than 10\,K, we set the
cooling rate to zero.  



Similar as in \citet{Joung_06}, we make a division into three temperature
regimes: the {\it Cold Neutral Medium (CNM)} up to 175\,K, where the first stable
part of the cooling curve ends, the {\it Warm Neutral Medium (WNM)} in the
intermediate temperature regime up to the ionisation temperature of hydrogen
and the {\it Hot Ionised Medium (HIM)} for higher temperatures (compare to
Fig.~\ref{fig:cool_curve}).

\subsection{Parameters of our standard model}
\label{sec:standard_param}

The parameters we use for our standard model are summarised in
Table\,\ref{tab:stanpar}. 
In the past subsections, we already motivated the supernova as well as the mass
injection rate. For the remaining parameters, 
short explanations are given in this paragraph.

The underlying gravitational potential is made up of two components:
\begin{dingautolist}{172}
  \item The potential of the nuclear star cluster, 
  which is modelled according to a Plummer distribution:
  \begin{eqnarray}
    \phi_{\mathrm{Plummer}}(r) = - \frac{G \, M_*}{\sqrt{r^2+R_{\mathrm{c}}^2}}. 
  \end{eqnarray} 
    \item The {\sc Newtonian} point-like potential of the nuclear black hole:
     \begin{eqnarray}
         \phi_{\mathrm{BH}}(r) = -\frac{G \, M_{\mathrm{BH}}}{r}.
     \end{eqnarray}
\end{dingautolist}

\begin{table}
\begin{center}
\caption[Parameters of our standard model]{Parameters of our standard model.}
 \label{tab:stanpar}
\begin{tabular}{cc|cc}
\hline
Parameter & Value & Parameter & Value \\
\hline
$M_{\mathrm{BH}}$ & $6.6\,\cdot 10^{7}\,M_{\sun}$ & $\alpha$ & 0.5 \\
$M_{*}$ & $1.9\,\cdot 10^{9}\,M_{\sun}$ & $T_{\mathrm{ini}}$ & $2.0\,\cdot 10^{6}\,$K \\
$M_{\mathrm{gas}}^{\mathrm{ini}}$ & $1.2\,\cdot 10^{4}\,M_{\sun}$ & 
$\dot{M}_{\mathrm{n}}$ & $6.0\,\cdot 10^{-9}\,M_{\sun}/(yr\,M_{\sun})$ \\
$R_{\mathrm{c}}$ & 25\,pc & $M_{\mathrm{PN}}$ & $2.2\,M_{\sun}$ \\
$R_{\mathrm{T}}$ & 5\,pc & SNR & $10^{-10}\, \mathrm{SNe}/(\mathrm{yr}\,M_{\sun})$ \\
$R_{\mathrm{out}}$ & 50\,pc & $\Gamma$ & $5/3$ \\
$\sigma_{*}$ & 165\,km/s \\
\hline
\end{tabular}
\end{center}

\medskip
 Mass of the black hole 
 ($M_{\mathrm{BH}}$), normalisation constant of the stellar potential ($M_{*}$), 
 initial gas mass ($M_{\mathrm{gas}}^{\mathrm{ini}}$), cluster core 
 radius ($R_{\mathrm{c}}$), torus radius ($R_{\mathrm{T}}$), outer radius ($R_{\mathrm{out}}$),
 stellar velocity dispersion ($\sigma_{*}$), exponent of the angular momentum distribution ($\alpha$),
 initial gas temperature ($T_{\mathrm{ini}}$), normalised mass injection rate ($\dot{M}_{\mathrm{n}}$), 
 mass of a single injection ($M_{\mathrm{PN}}$), 
 supernova rate (SNR) and adiabatic exponent ($\Gamma$).
\end{table}

A core radius $R_{\mathrm{c}}$=25\,pc, a black hole mass $M_{\mathrm{BH}}=6.6\,\cdot 10^{7}\,M_{\sun}$
and a normalisation factor for the stellar distribution of $M_{*}=1.9\,\cdot 10^{9}\,M_{\sun}$ is used. 
The latter leads to a total integrated stellar mass within the core radius of $6.7 \cdot 10^8\,M_{\sun}$.
Our initial condition comprises of a {\it TTM}-model similar to the one
discussed in \citet{Schartmann_05}, but with a Plummer potential
(as given above) and the
parameters listed in Table\,\ref{tab:stanpar}.  
The angular momentum of the stellar content, 
the gas in the initial condition as well as
the injected gas is given by the following distribution:
\begin{eqnarray}
  j_{\mathrm{spec}} = \sqrt{G \, (M_{\mathrm{BH}} \, + M_*(R_{\mathrm{T}})) \,
 R_{\mathrm{T}}}
 \, \left( \frac{R}{R_{\mathrm{T}}}  \right)^{\alpha}.
\end{eqnarray}
with a constant $\alpha=0.5$ and a torus
radius of $R_{\mathrm{T}}=5\,pc$, where
the torus radius is defined as the minimum of the effective potential.
The turbulent pressure of the TTM-model 
is substituted by thermal
pressure. For the case of our standard model, the velocity dispersion of the
assumed dust clouds in the {\it TTM}-model of $\sigma_*=165\,$km/s 
then corresponds to an initial
temperature of $T_{\mathrm{ini}}$ = $2.0\,\cdot 10^{6}\,$K.  
The mass loss rate was chosen to be $6.0\,\cdot
10^{-9}\,M_{\sun}/(yr\,M_{\sun})$, which means a mass injection of $6.0\,\cdot
10^{-9}$ solar masses of gas per year, normalised to one solar mass of
stars (see discussion in Sect.~\ref{sec:mass_input}). A SN-rate of
$10^{-10}\,$ supernovae per year and normalised to $1\,M_{\sun}$ in stellar mass 
was used
(see Sect.~\ref{sec:en_input} for further discussion).
We apply the equation of ideal
gas as a closing condition with an adiabatic index of an ideal monoatomic gas ($\Gamma =
\frac{5}{3}$). The gas has solar abundances in our simulations, with
a mean molecular weight, which changes from 0.6 (fully ionised
gas with solar abundances) to 1.2, following a smooth transition
at the ionisation temperature of hydrogen.
Most of the simulations discussed in this paper ran for an evolutionary period of
10\,orbits at the torus radius ({\it global orbits}\footnote{Global orbit always refers to 
a Keplerian orbit, measured at the torus radius at $R_{\mathrm{T}}=5\,$pc and corresponds
to a real time of approximately $1.2\cdot 10^5$\,yr.}), 
corresponding to $1.2\,\cdot 10^6\,$yr. 
Note that this is a tiny fraction of the age of the stellar population.
Therefore, evolutionary effects can be ignored.

\section{Numerical method}
\label{sec:numerics}

For our simulations, we use the 3D radiation hydrodynamics code {\sc TRAMP}
\citep{Kley_89,Klahr_98,Klahr_99}, which uses similar numerical schemes as
the Zeus code \citep{Stone_92a,Stone_92b,Stone_92c}. 
It solves the equations of ideal hydrodynamics with the help of an
{\it Operator Splitting} technique on a spherical grid that is 
logarithmically spaced in the radial direction. 
Using this method, one has always an optimal resolution and at the 
same time more or less boxy e.~g.~equi-spaced grid cells. 
For the advection step, the monotonic transport scheme of \citet{Leer_77} is applied,
which is a spatially second-order-accurate upwind method. The radiation
transport in {\sc TRAMP} is treated in the flux-limited diffusion approximation 
using the flux limiters of \citet{Levermore_81}.  
Additional source terms for mass input (Sect.~\ref{sec:mass_input}) and 
energy input (Sect.~\ref{sec:en_input}) are treated in a separate step. 
{\sc TRAMP} discretises the hydrodynamic equations in the
finite volume description and, therefore,
locally and globally conserves advected quantities. 
The force step is calculated 
using a finite difference discretisation. 
Shocks are treated by introducing
a non-linear artificial viscosity, 
which smears out discontinuities over several
cells and, thereby, results in the correct generation of entropy in the shock,
the correct shock velocities and suppresses post-shock oscillations.
It is treated as an additional contribution to the pressure.
The non-linearity ensures that the artificial viscosity is large in shocks, but negligible
elsewhere.  

\subsection{Domain decomposition and boundary conditions}
\label{sec:domaindeco_boun}

In order to gain resolution,  
we split the whole computational domain into
three parts: the inner one ranges from $0.2\,$ to $2.0\,$pc, the middle
one from $1.0\,$ to $10.0\,$pc and the outer part from $5.0\,$ to
$50.0\,$pc (logarithmic grid). Towards the centre, the size of the single cells decreases
and, additionally, the obtained velocities increase due to
the presence of the central gravitational potential and the conservation of
angular momentum of infalling material. This leads to ever smaller timesteps, 
in order to prevent information to be transported along more than one cell during a single
timestep ({\it Courant-Friedrichs-Levy} condition, \citealp{Courant_48,Ritchmyer_67}). 
Hence, we are able to calculate a
much longer time period for the outer domain with the same amount of CPU usage
and the same grid size, compared to the inner domain. On the other hand,
further in, processes act on much shorter time-scales and only a shorter time
evolution is needed in order to obtain a steady state. 
Therefore, we calculate 10\,global orbits 
for the outer domain, but only the last orbit for the middle domain and one
tenth of an orbit for the innermost part. This leads to a comparable amount of
CPU time of approximately one weak per domain. 

In the first step, the outer domain is calculated. The physical state of
material at the position of 
the outer boundary of the middle model is stored in a file every 0.1\% of a global orbit.
Outflow boundary conditions are
used radially inwards as well as outwards. 
This means we allow for the outflow of matter from the grid and minimize 
the reflection of waves at the grid edge, while not allowing for inflow, 
thus preventing the boundary cells from generating a numerical 
instability. 
During the calculation of the
middle domain in the second step, the inner radial boundary is set to outflow,
whereas the outer boundary is set by the afore stored quantities\footnote{A
linear interpolation in time is used.}, which can
result in in- or outflowing material, thanks to the wide overlap of the domains. 
At the same time, physical quantities advected across the position of the outer boundary of
the inner domain are stored. In the last step, the same is done for the
innermost model space. 
For the final quantitative analysis and the subsequent radiative transfer calculations, the three
domains are combined again. 
This procedure is a reasonable approximation, as in most of the simulations we observe infall
of gas towards the inner region. Outflowing material can partly be taken into
account, as the domains have a quite large overlapping region.   

Concerning the other coordinates,  we model a range of $90\degr$ with 
periodic boundary conditions in $\phi$
direction.
In $\theta$-direction, where we model from 4 degrees to 176 degrees, 
as well as for the inner- and outermost boundary in radial direction, outflow boundary
conditions are applied in a sense that mass is allowed to flow out
of the integration domain, but no material can get back into it. 
This means that the values of the {\it ghost cells} 
\footnote{{\it Ghost cells} are additional boundary cells, not belonging to the simulated
physical domain, but necessary to implement boundary conditions.} are set such, that
the normal derivatives of the boundary values equal to zero for all quantities, 
presenting a smooth continuation of the flow through the boundary.
Furthermore, we only allow outward directed motion, otherwise the normal velocity components
are set to zero. 
For each of the domains, we are able to calculate 75 cells in radial, 94 cells in $\theta$ and 
49 cells in $\phi$ direction, leading to cell sizes between 0.006\,pc in the innermost part of 
the whole domain and 1.5\,pc at the outer edge. 

\subsection{Initial condition}
\label{sec:ini_con}
A hydrodynamically stable
initial condition for the case of our effective potential -- made up of the
nuclear stellar cluster, the black hole and the additional centrifugal potential
due to orbital rotation -- is given by the isothermal {\it Turbulent Torus Model}
({\it TTM}, \citealp{Camenzind_95}, \citealp{Schartmann_05}).  
A conical volume around the rotation axis with a half opening angle of
$4\degr$ needs to be cut out, because the
funnel region of this field configuration
is dynamically unstable and would cause unphysical behaviour of the state vector.

\section{Results}
\label{sec:results}


\begin{figure*}
  \centering
  \resizebox{0.95\hsize}{!}{\includegraphics{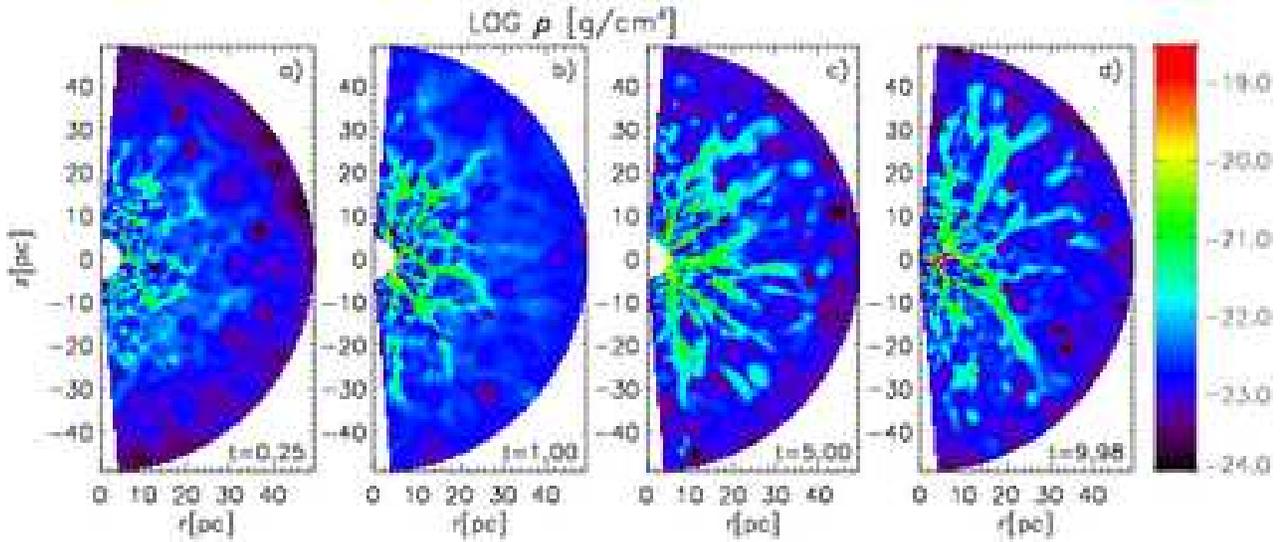}} 
 \caption[Temporal evolution of the density in our standard
    model]{Temporal 
    evolution of the density in our standard model, given in a
    meridional slice of the torus. Shown are four stages of the torus at
    0.25, 1.0, 5.0  and roughly 10.0\,global orbits. The scaling is logarithmic.} 
  \label{fig:time_series_a007_new.eps} 
\end{figure*}

\begin{figure*}
  \centering
  \resizebox{0.75\hsize}{!}{\includegraphics{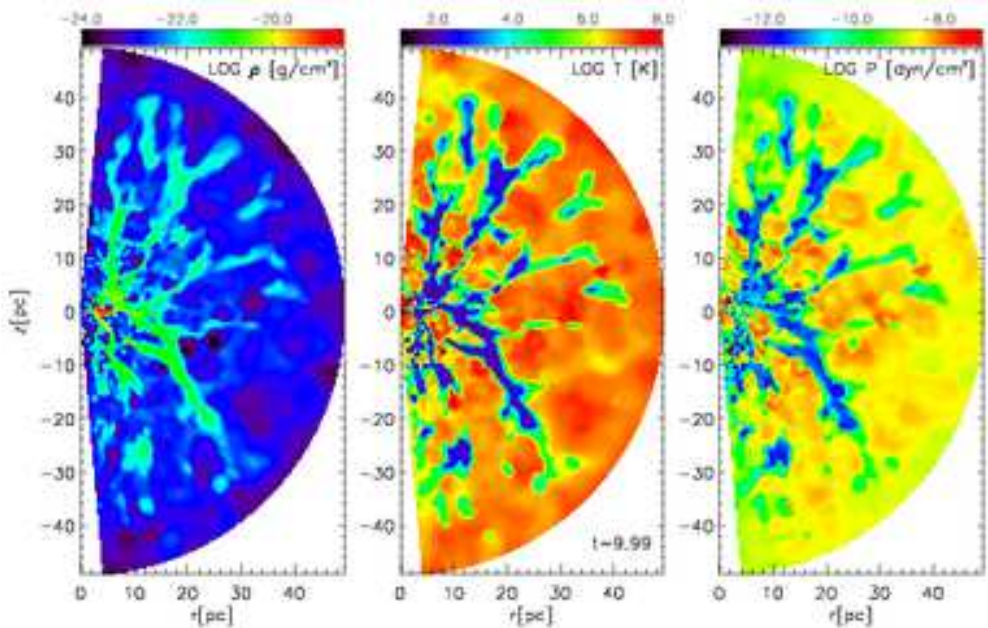}} 
  \caption[State of the torus after roughly ten global orbits, meridional plane]{State of the torus
    after roughly ten global orbits. Shown are from left to
    right the density distribution, the temperature distribution and the
    pressure distribution of a meridional slice through the 3D data cube. 
    All images are displayed in logarithmic scaling.} 
  \label{fig:a007_tsli_999.eps} 
\end{figure*}

\subsection{The evolution of our standard model}
\label{sec:torus_evolution}

Within the first few percent of a global orbit, single injected blobs of gas are
visible, mainly within the central 25\,pc, the core radius of the stellar
distribution (Fig.~\ref{fig:time_series_a007_new.eps}a). Spherical cavities 
are formed due to the explosion of
supernovae. While these explosions predominantly form shock fronts, the planetary
nebulae interact in two different ways, as we could see in separate simulations of
single events (see \citealp{Schartmann_07a}): In the subsonic case (e.g.\/ hot surrounding medium), colliding
streams tend to stick together and form long filaments in perpendicular
direction from the initial direction. The supersonic interaction of two
planetary nebulae (e.g.\/ in cold medium) will cause the formation of a common
shock front, which surrounds the point of interaction.
With ongoing evolution, the single blobs collide to create a filamentary,
net-like structure, enhanced and further shaped by supernova explosions 
(Fig.~\ref{fig:time_series_a007_new.eps}b and c).  
This process is also amplified by the onset of cooling instability in the dense regions
of the filaments.
Cooling also leads to loss of thermal pressure and, therefore, material tends to
fall towards the minimum of the potential well, located at the torus radius (5\,pc)
within the equatorial plane of the torus. Around this radius, the filamentary
structure opens out into a geometrically thin, but optically thick and very 
turbulent disk (Fig.~\ref{fig:time_series_a007_new.eps}d).   

This process is comparable to the so-called galactic fountain process
\citep{Shapiro_76} visible in galactic disks. The more tenuous the gas, the more
effective are the supernova explosions in pushing away material. 
This mechanism finally leads to the large-scale, filamentary structure
visible in Fig.~\ref{fig:time_series_a007_new.eps}d.  

As the implemented optically thin cooling scales proportional to the gas density
squared, only the dense filaments cool on short time-scales, whereas the
interclump-medium is effectively heated by the energy injection of the
supernova explosions. Accordingly, the gas
velocity can appreciably differ in various temperature regimes, as will be
discussed in Sect.~\ref{sec:dyn_state}.  

\begin{figure*}
  \centering
  \resizebox{0.75\hsize}{!}{\includegraphics{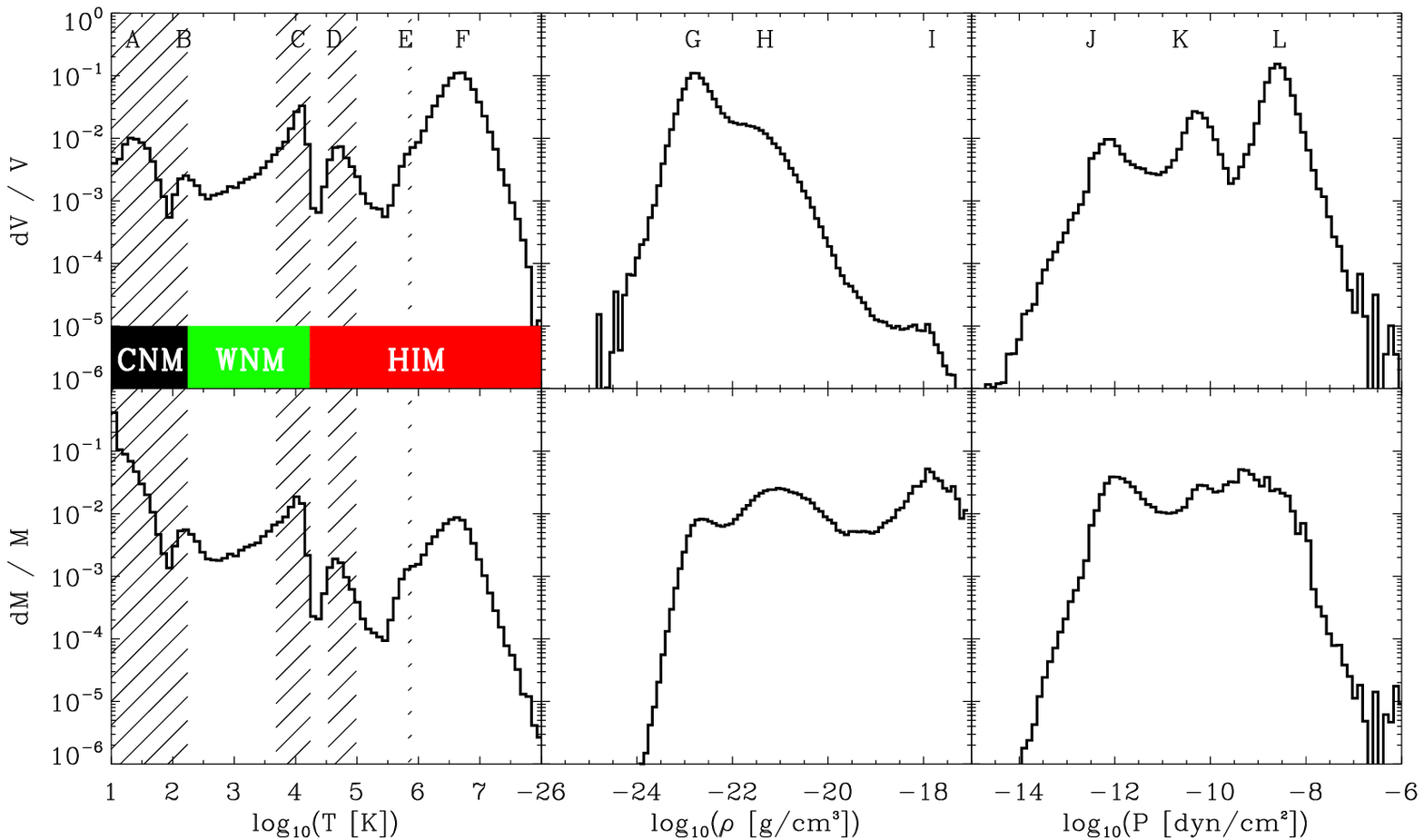}} 
  \caption[Phase diagrams for the total model space]{Phase diagrams for the total
    model space of our standard model 
    after an evolution of ten global orbits. The hatched regions indicate
    thermally stable regimes according to the cooling curve (see
    thick red line segments in Fig.~\ref{fig:cool_curve}).} 
  \label{fig:a007_total_phase} 
\end{figure*}
 
In order to ensure that the memory of the initial condition is completely lost, 
we start calculating the middle part of the computational
domain after an evolution of nine global orbits of the outer domain. 
Already after a fraction of the simulated one global orbit, a
geometrically relatively thin, but optically very dense disk forms in the
equatorial plane in the vicinity of the minimum of the potential  
(see Fig.~\ref{fig:a007_turbdisc_paper.eps} and discussion in Sect.~\ref{sec:turb_disk}). 

The final state of our standard model after ten global orbits is depicted in
terms of density, temperature and pressure within a meridional plane in
Fig.~\ref{fig:a007_tsli_999.eps}. A multiphase medium has formed
(Sect.~\ref{sec:multiphase}). The temperature and the density
distribution are complementary in the sense that high density regions possess
low temperature due to rapid cooling and hot regions possess low density as
material has been blown away by supernova explosions. It is important to note that the pressure
distribution (right panel in Fig.~\ref{fig:a007_tsli_999.eps}) is far away from pressure
equilibrium, as it spans more than five orders of
magnitude.

\subsection{Multiphase medium}
\label{sec:multiphase}

Fig.~\ref{fig:a007_total_phase} shows phase diagrams for the total
model space of our standard model (all three spatial subregimes) after ten global orbits. The hatched
regions indicate thermally stable parts of the cooling curve (see
Sect.~\ref{sec:model}). Capital letters (A-F) denote the six bumps in the
probability density function for gas temperature (upper left panel). 
C and D as well as the small
excess E can be directly related to the stable regions of the cooling curve,
because time spent in each temperature range is proportional to the inverse of
the cooling function \citep{Gerola_74}. At these temperatures, gas cools
comparatively slow and therefore assembles in these regimes. The explanation
for the maximum B is identical. The minimum in-between A and B evolves,
because gas at these temperatures is mostly located in dense clumps. As
cooling depends on the square of the density, it can cool quite fast, even if the
cooling curve indicates thermally stable behaviour. Peak F can be explained by
constant energy input from supernova explosions, which allows gas to reside even
in the thermally unstable regime (see also discussion in \citealp{Avillez_05}). 
Looking at the mass distribution of the various temperature phases
(lower left panel), a similar curve is obtained, but tilted, as most of the mass is
concentrated in the cold temperature regime (dense filaments and disk), 
while the hot gas fills most of the volume (supernova blown cavities).

Plotted with respect to the gas density (second column of
Fig.~\ref{fig:a007_total_phase}), three local maxima are visible
(G,H,I). 
Feature I is caused by the very dense disk component, as it is dominant in
mass, but unimportant in volume. This material had enough time to cool to
temperatures below 175\,K and, therefore, corresponds to maximum A and B in
the left panel. G can be accounted to regions, where supernova explosions
happened recently, contributing to a large volume fraction, but only little to
the total mass. In the left column, this corresponds to feature F. The bump
(H) in between arises from material in the large scale filaments, which is in
the process of settling down and thereby cooling. 

The distribution of mass and volume with respect to pressure is shown in the
right column of Fig.~\ref{fig:a007_total_phase}. As already
evident from the two-dimensional cuts in Fig.~\ref{fig:a007_tsli_999.eps}, 
no pressure equilibrium is reached. This fact is in qualitative agreement with recent
observations of \citet{Jenkins_07} in our local neighbourhood, 
who find variations in the pressure of the
diffuse, cold neutral medium in a range of 
$10^3$\,K$\,\mathrm{cm}^{-3} < P/\mathrm{k}_{\mathrm{B}} < 10^4\,$K\,$\mathrm{cm}^{-3}$
by using C\,{\sc i} absorption lines, measured in the ultraviolet
spectra of roughly 100 hot stars, taken from the HST archive. 
Instead of a single and sharp peak (expected for pressure equilibrium), we obtain
three local maxima (right panel in
Fig.~\ref{fig:a007_total_phase}). 
The system tries to reach an equilibrium value (located around peak K). 
Peak L can be assigned to
regions with recent supernova action, leading to the highest pressure values
of the simulation and filling a large volume. This is the same component
causing the peaks F and G. We interpret deviations from a global
equilibrium pressure value towards small pressure (peak J) as a sign of ongoing
thermal instability within the fast cooling, densest regions of the flow (see
also discussion in \citealp{Joung_06}). This interpretation is backed by
Fig.~\ref{fig:a007_tsli_999.eps}, where cold and dense regions correspond to low pressure states
and hot and tenuous regions to high pressure values.  

In terms of volume and mass filling factors, the following picture
evolves: Most important in mass is the CNM with 60.4\% of the total mass,
mostly residing in the disk component and the inner parts of the long
filament, which had enough
time to cool to temperatures of the CNM and occupies the smallest fraction in
volume (only 5.4\%). Second most important (22.6\% in
mass) is the WNM, made up of material in the process of cooling down. Only
16.9\% of the mass is in the HIM, which occupies most of the space (83.3\% of
the volume). Only 11.3\% of the whole model space is occupied by the WNM.

\subsection{Dynamical state of the torus}
\label{sec:dyn_state}

\begin{figure}
  \centering
  \resizebox{\hsize}{!}{\includegraphics{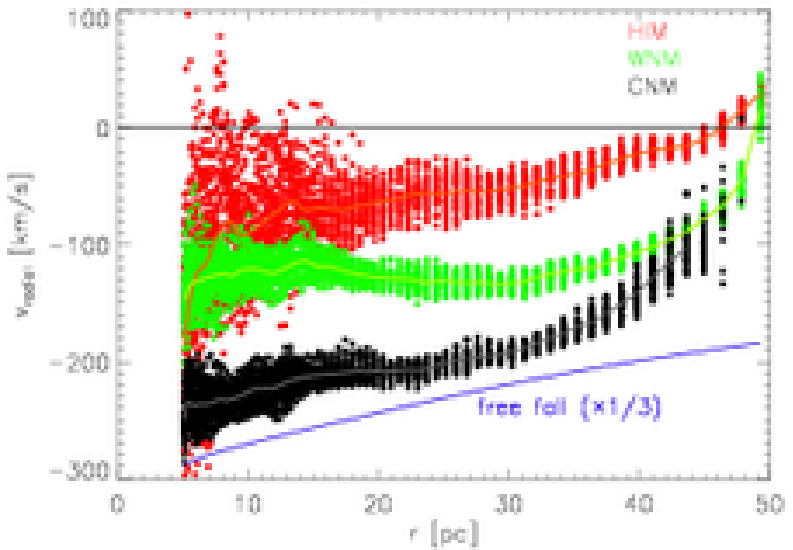}} 
  \caption[Radial velocities of the outer model space between eight and ten global orbits, separated into temperature
    phases]{Radial velocities of the outer model space between eight and ten global orbits, separated into temperature
    phases. Negative velocities refer to infalling motions. 
    The blue line represents the free fall velocity ($v_{\mathrm{ff}}$), scaled with a
    factor of $1/3$, the dashed line symbolises the location of the core
    radius of the stellar distribution.} 
  \label{fig:rad_time_out_a007.eps} 
\end{figure}

\begin{figure*}
  \centering
  \resizebox{0.75\hsize}{!}{\includegraphics[angle=0]{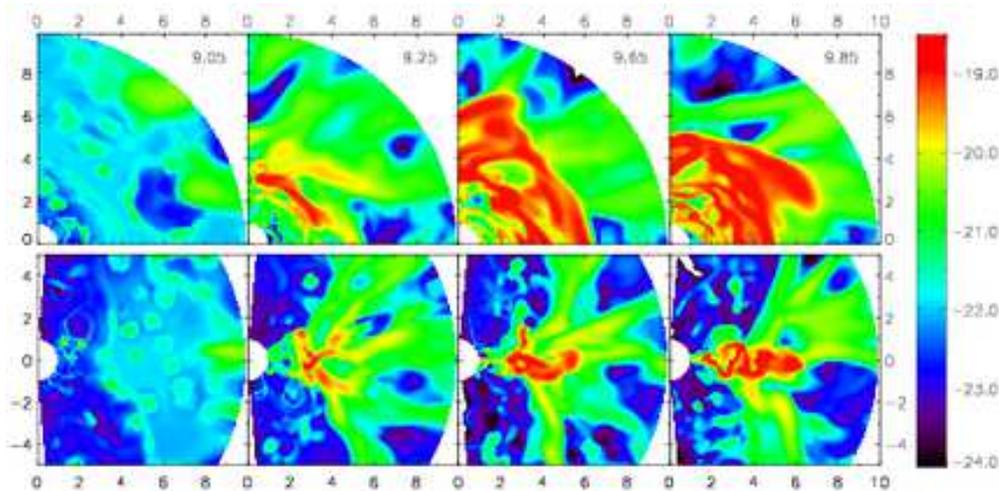}} 
  \caption[Formation of a turbulent, geometrically thin disk.]{Formation of a turbulent, geometrically thin disk 
   close to the midplane. Shown is the logarithm of the density in units of
   g/$\mathrm{cm}^3$ as a cut through the midplane (upper row) and through the central meridional plane (lower
   row). Time is given in global orbits. Labels are given in pc.}  
  \label{fig:a007_turbdisc_paper.eps}
\end{figure*}

The build-up of the disk and torus component can also be quantified
by means of radial velocity distributions, as given in
Fig.~\ref{fig:rad_time_out_a007.eps}. The velocity is averaged over all angles $\theta$
and $\phi$ and separated into the three temperature components. For an
integration time between eight and ten global orbits, the data are plotted
every 0.05 global orbits (single dots). The solid lines are temporal averages 
for the whole range of displayed data. For comparison, we show 
the free fall velocity of a test
particle within the total gravitational potential of the black hole and the
stellar distribution, scaled down by a factor of three. 
It is a remarkable feature of our simulations that the slope of
the velocity distributions of all three components are similar to {\it free fall}.

Clearly, the hot tenuous medium is able
to provide the largest thermal pressure support, which slows down the infall
towards the mass centre. In contrast to this, cold material sinks 
much faster to the bottom of the potential well, within the long, dense 
filaments. 

At the border
line between the phases with different flow structures, Kelvin-Helmholtz and
Rayleigh-Taylor instabilities are expected to occur, but are not visible in
our simulations due to the lack of sufficient resolution. They are supposed to increase
turbulence and promote further mixing. 
Due to their high density, the filaments are very stable and single supernova
explosions are not able to disrupt them.
In them, the CNM reaches radial velocities of almost 30\% of the free fall velocity, the
WNM approximately 15\% and the HIM 10\%. 
This is a further indication that energy in form of supernovae or due to 
the planetary nebulae injection is mainly dispensed into the hot medium, which is 
also visible in the larger scatter of the radial velocities of the hot component.

\subsection{Formation of a turbulent disk}
\label{sec:turb_disk}

As already discussed in Sect.~\ref{sec:torus_evolution}, in the vicinity of the torus radius, a
geometrically rather thin, but optically very dense disk forms. 
Fig.~\ref{fig:a007_turbdisc_paper.eps} shows
the temporal evolution of density for four snap-shots of the middle domain.
Its dynamical evolution is mainly
determined by mass inflow through the outer radial boundary. This material
reaches the middle part both in the form of broad streams as well as in blobs and
filaments of material and gets structured by supernova explosions and
planetary nebulae injection. As more and more material falls towards the minimum of the
potential, a turbulent, fluffy disk emerges between the inner boundary and
roughly six or seven parsec. We note the remarkably filamentary structure of the
disk, showing strings of material, elongated in azimuthal direction (for 
reasons identical to those already discussed for the outer domain). 
The difference is that the average flow is here directed in 
azimuthal direction and not in radial direction.

Altogether, the evolution
can be summarised as a collapse from a geometrically thick distribution
(the {\it torus} component) to a
geometrically thin disk. The initial
torus is mainly (thermal) pressure supported and,
therefore, collapses due to radiative cooling processes into dense
filaments, as cooling and heating of the gas is locally unbalanced. 
This finally leads to the formation of the 
thin disk in the vicinity of
the minimum of the gravitational potential, close to the angular momentum
barrier. Thermal pressure does not play a
role anymore, as the disk has cooled to temperatures of less than 100\,K and reaches
in some regions even our minimal temperature of 10\,K. The scale height is now
determined by random motions of the filamentary disk 
and rotation. This hydrodynamical turbulence leads to
angular momentum transfer outwards and accretion towards the centre.

\subsection{Parameter studies}

In order to investigate the dependencies of our model, 
we conducted several parameter studies, which are briefly presented 
here (see \citealp{Schartmann_07a} for a more detailed discussion). 

\begin{dingautolist}{172}
\item {\it Mass and energy input rate:}
Changing the mass injection rate and the supernova rate, we found that a
complementary behaviour is obtained. Both, with
an increasing supernova rate or a decreasing mass-loss rate, the inward bound
flow is more and more turned into an outflow, as demonstrated for example in 
Fig.~\ref{fig:rad_vrad_out_a00567_HIM.eps}.
This has also been shown in 1D hydrodynamical simulations of the interstellar medium in 
elliptical galaxies by \citet{Gaibler_05}. 
The transition radius between inflowing and outflowing hot gas increases with increasing 
mass loss rate or decreasing supernova rate. Inside the transition radius, the curves flatten for 
models with larger mass input rate. This is caused by the additional turbulent pressure 
component, provided by the PNe injection of gas blobs with random velocities. 

\begin{figure}
  \centering
  \resizebox{\hsize}{!}{\includegraphics{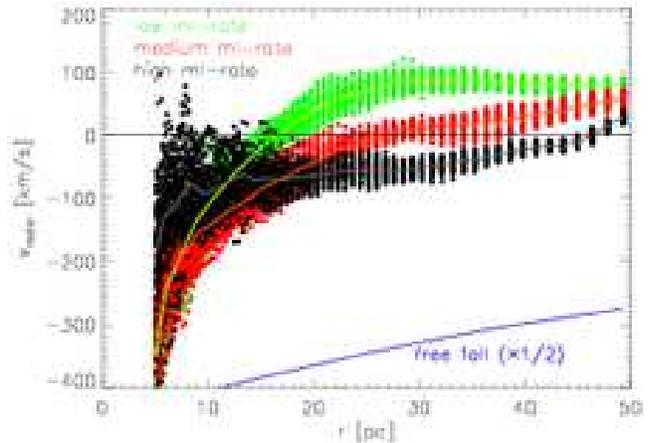}} 
  \caption[Comparison of radial velocities within the mass loss rate study for
           the HIM between global orbit eight and ten]{Comparison of radial velocities within the mass loss rate study for
           the HIM between global orbit eight and ten. Lines represent time-averaged
           (global orbits eight to ten) values.} 
  \label{fig:rad_vrad_out_a00567_HIM.eps}
\end{figure}

\item{\it Cooling rate:}
Cooling is a sensitive function of metallicity. The larger the metallicity, the larger the 
cooling rate. We varied the cooling rate to account for the unknown 
metallicity of the gas in AGN tori. The most striking difference appears in
the broadening of the filaments, when increasing the cooling rate, as gas of even
lower densities is able to effectively cool to CNM temperatures.  

\item{\it Mass of the central stellar cluster:}
Varying the (generally unknown) mass content in stars is of interest, 
as it first determines 
the gravitational potential in the outer part beyond the influence of the black hole. 
Second, it changes the amount of 
supernova input and mass input in equal measure. 
A higher stellar mass produces a deeper potential well and therefore accelerates
the gas to higher velocities and the transition radius between inflow and outflow of 
the hot temperature phase increases. For the case of a higher total stellar mass, 
the evolving structures are more compact and a more massive disk
is able to form at the angular momentum barrier.  

\item{\it Various initial conditions:}
In order to check the dependence of our results on the initial condition,
we started one simulation with an initially empty model space 
(possessing a homogeneous density distribution with the lower density threshold of our 
simulations -- $10^{-26}$\,g/$\mathrm{cm}^3$). As expected, this has no visible consequence 
on the evolution of the gaseous structures and a steady state in terms of total gas mass 
is already reached after less than 3 global orbits.

Also starting with cold ($T\,=\,1000$\,K) initial conditions, where the initial torus is 
supported by turbulent 
pressure on small scales yields identical results.  
This small-scale turbulence decays within a fraction of an orbit and is dissipated into heat. 
Additionally, the supernova input quickly heats the tenuous medium outside-in towards the 
minimum temperature at the torus radius. Therefore, after a short time, we again get a very 
similar behaviour as in our standard model. 

\end{dingautolist}

\section{Discussion}
\label{sec:discussion}

It needs to be emphasized that within our model, the large-scale torus component 
is not a long-living steady state,
but it can only exist as long as mass and energy is injected from the stellar cluster.
After switching off energy and mass input,  
most of the large scale structure will disappear within approximately 1.5 global orbits.
The turbulent disk in the vicinity of the angular momentum barrier is the only remaining structure. 
In reality, this shutdown will happen naturally due to the evolution of the nuclear star 
cluster, but on much longer time-scales.  
It means that -- within our modelling -- the existence of tori is closely linked to
the existence and evolution of a stellar cluster in the centre of a galaxy. This fits well
to the observations by \citet{Davies_07}, which reveal a 
link between nuclear starbursts and AGN activity with a certain time delay in-between. 
Furthermore, it needs to be stressed that the torus scale height in our modelling is mainly
determined by the spatial distribution of the stars. 

The total simulated time of our runs is of the order of 10 global orbits, which corresponds to
$1.2$\,Myr. This is
sufficiently long to avoid memory effects from the initial condition and at
that point, a global dynamical equilibrium concerning the total mass and 
total energy is reached for the outer domain. In the middle and inner domain,
the calculated time evolution of one and 0.1 global orbits
is not long enough to reach a global dynamical
equilibrium state. However, the technical limitation of this study
did not allow a longer evolution.

\subsection{Limitations of our simulations and future work}
\label{sec:limits}

This paper presents an exploratory study of a self-consistent hydrodynamical
model of gaseous and dusty AGN tori. 
Due to the simplifications required by the efficiency of our code, some 
important problems and astrophysical effects remain open:

\begin{dingautolist}{172}
\item {\it Longer time evolution:} With our serial code we are limited
  in temporal evolution to about 1.2\,Myrs, while typical
  evolutionary times of such systems are of the order of 10 to
  100\,Myrs. Therefore, we currently port our routines to the parallel {\sc PLUTO}
  code \citep{Mignone_07}.
\item {\it Radiation contribution from central source and stars:} 
  The effect of the radiation of the central
  accretion disk is twofold: First of all, it contributes to the heating of
  the dust and gas. Second, its radiation pressure plays an important role in the dynamics of
  AGN dust tori, as was recently shown by \citet{Krolik_07}. 
  With the help of our detailed radiative transfer calculations, we compared
  the accelerations due to radiation pressure of the dust reemission with the 
  gravitational accelerations in vertical direction for each cell within the
  computational domain of the snapshot after roughly 10 global orbits, using a flux mean dust opacity 
  in combination with a grey approximation. In this estimation, infrared radiation pressure significantly contributes only 
  close to the midplane in the inner few parsecs from the heating source, which is in accordance with the detailed 
  calculations of \citet{Krolik_07}. Within the large scale filaments, the ratio between the accelerations due to the infrared radiation 
  pressure and the gravitational acceleration in vertical direction amounts to
  a few percent only, except for the direct illuminated parts of the
  filaments, where it can also exceed the gravitational accelerations and
  will therefore drive an outflow along the torus axis, 
  widening the (so far too narrow) opening angle in our simulations. However,
  detailed radiative transfer calculations during the whole course of the
  hydrodynamical simulations are necessary to finally clarify this question.
  Opposed to \citet{Thompson_05}, who study radiation pressure-supported
  starburst disks, we are currently interested in the 
  phase following an efficient starburst. By then, the stellar luminosity has already
  dropped by a large factor. Giving the low stellar luminosity densities
  derived from the observations of a Seyfert galaxy sample discussed in
  \citet{Davies_07}, the stellar luminosity seems to be only a small fraction
  of the luminosity of the central accretion disk 
  at this evolutionary stage.
  Therefore, we currently neglect its
  contribution to the dynamics, as well as to the heating of the dust within 
  our postprocessing radiative transfer calculations. 
\item {\it Magnetic fields:} Weak magnetic fields in combination with shear
  due to differential rotation cause magneto-rotational instability
  \citep{Balbus_91}, which provides a means to transport angular momentum and thus  
  enables accretion towards the core (that is through the inner boundary of our  
  computational domain).
  Probably coexisting strong magnetic fields in a spatially
  distinct region cause hydromagnetic outflows, which might be
  preferentially directed along the torus axis and thereby widening the
  opening angle of the torus.
\item {\it Self gravity and star formation:}
  At the current state of our modelling, self-gravity of the gas is not taken into
  account. However, examining the Jeans-criterion for each cell of our model space, we find 
  that part of the inner turbulent disk, as well as a small fraction of the 
  filamentary structure is Jeans unstable and will form stars. 
  This is consistent with the findings 
  of \citet{Nayakshin_06a} and \citet{Paumard_06}. 
  In the snapshot after 10 orbits, the unstable mass amounts to 0.3\% in
  volume or 15\% in mass respectively. 
  However, it should be emphasized again that our purely hydro- 
  dynamical code is not capable to treat
  the accretion flow towards the core correctly, leading to a pile-up of  
  cold material in the disk which does
  not describe the steady state density.  
  Subsequently, supernova type~II explosions might 
  inflate or even partly 
  disrupt the evolving dense and geometrically thin turbulent disk. 
  However, one should note again that 
  detailed simulations
  including self gravity will be subject of our future work.   
\item {\it Further mass input:} There will also be material transported into 
  the simulated domain of the active galactic
  nucleus from further out of the galaxy. Prominent examples
  are dusty mini-spirals, of which some have been discovered by NACO 
  \citep{Prieto_05} or HST-ACS observations (Hubble Space Telescope -- Advanced
  Camera for Surveys). 
  Simulations \citep{Maciejewski_06} show that with the help of such nuclear
  spirals, enough gas inflow can be produced in order to power luminous nearby
  AGN.

\begin{figure*}
  \centering
  \resizebox{0.9\hsize}{!}{\includegraphics[angle=0]{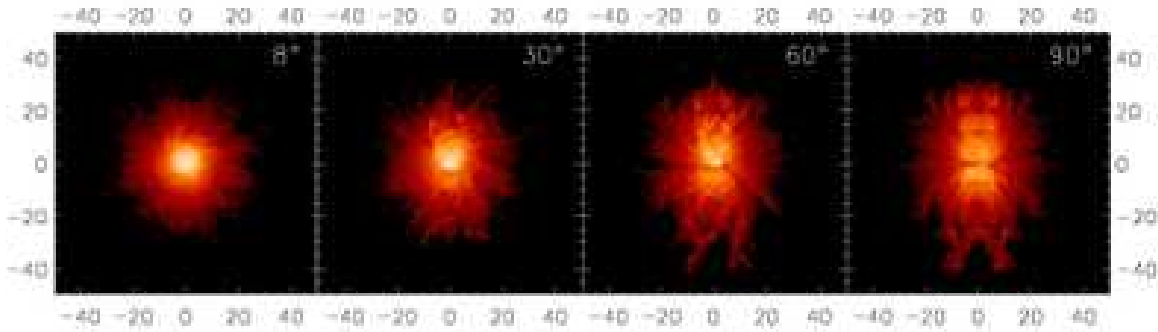}} 
  \caption[Surface brightness distributions of our hydrodynamic standard
           model]{Images of our 
           standard model at $12\,\umu$m. Shown are the inclination angles
           $8\hbox{$^\circ$}$, $30\hbox{$^\circ$}$, $60\hbox{$^\circ$}$ and $90\hbox{$^\circ$}$ 
           for a common azimuthal angle of $\phi=45\hbox{$^\circ$}$ with a logarithmic colour scale.
           Labels are given in pc.} 
  \label{fig:mc3d_hydro} 
\end{figure*}
  
\item {\it Detailed data comparison:}
  After including part of the discussed additional features, one should do 
  a detailed comparison with observational data. Starting from the simulated 
  surface brightness distributions, we are able to 
  calculate visibilities in order to compare to recent mid-infrared interferometric
  observations. 
  Furthermore, SINFONI observations \citep[e.~g.~][]{Davies_07} of Seyfert nuclei 
  and starbursting galaxies will give us important constraints for our modelling, 
  e.~g.~in terms of the nuclear stellar cluster, 
  as well as fuelling of the central active region. 
\end{dingautolist}

\section{Comparison with observations}
\label{sec:comp_obs}

The only possibility to directly resolve the dust structure in nuclei of Seyfert galaxies 
so far is with the help of the MID-infrared Interferometric instrument MIDI at the VLTI. 
Thereby, tori were found in the closest Seyfert nuclei NGC~1068 \citep{Jaffe_04,Raban_08}
and Circinus \citep{Tristram_07}. Remarkably, both objects are well modelled by a warm, more
or less spherical component and a less extended, hotter elongated component, 
which might be interpreted as a disk seen edge-on.
The latter component coincides with an edge-on disk observed in maser emission 
(\citealp{Greenhill_03} for the case of the Circinus galaxy).
Some of these interpretations qualitatively match well with the results of our simulations
(see discussion in \citealp{Tristram_07}). 
It has to be taken into account that our modelling is meant to be suitable for an average 
Seyfert galaxy.

\subsection{Observing our standard model}
\label{sec:hydro_observable}

The basic procedure of our effort of simulating the observational properties
of dust tori is subdivided into several steps:
First of all, hydrodynamic simulations are carried out with the help of the
{\sc TRAMP} code as described above. 
In a postprocessing step, the resulting gas density distribution 
is converted into the corresponding dust density distribution with the help of a constant 
gas-to-dust-ratio of 250 applied to those patches of the gas
distribution, where temperatures are below the sublimation temperatures 
($T_{\mathrm{sub}}$) of the
various grain species. Our dust model takes graphite ($T_{\mathrm{sub}}^{\mathrm{graph}}\approx 1500\,$K) 
as well as silicate grains ($T_{\mathrm{sub}}^{\mathrm{sil}}\approx 1000\,$K) into account. 
This mixture leads to a prominent spectral feature at 9.7\,$\umu$m and a less pronounced 
feature at 18.5\,$\umu$m.
A slightly higher gas-to-dust ratio compared to the local galactic value of 160 
\citep{Sodroski_94} was applied, as we expect it to increase towards the centres of active
galaxies, as dust is likely to be destroyed in the harsh environment close to the
energy source. 
The resulting dust density distributions are then fed into  
the 3D radiative transfer code {\sc MC3D} \citep{Wolf_99a,Wolf_03}.
Here, in a two-step approach, first the temperature distribution is calculated, that is then used
to simulate observable quantities like spectral energy distributions (SEDs) or 
images at various wavelengths.
Grain-size dependent
sublimation (as described in \citealp{Schartmann_08}) is only considered during the radiative transfer
calculations and, therefore, only for dust in the close vicinity of the AGN. 
Here, we split the dust opacity model into 6 different grains, made up of the
smallest ($0.005\,\umu$m) and largest ($0.25\,\umu$m) grain of the MRN\footnote{The dust model 
is labelled according to the initials of Mathis, Rumpl and Nordsieck.}-size distribution
\citep{Mathis_77}
and the three grain species: astronomical silicate and the two orientations 
of graphite grains \citep{Draine_84,Laor_93,Weingartner_01}. 
As radiative transfer calculations are very time and memory consuming, the hydrodynamic grid 
-- an assembly of all three domains -- has to be mapped on to a slightly smaller grid. 
A resolution of 114 grid cells in radial direction (instead of 181 for the
composed hydrodynamic model), 61 grid cells in $\theta$ direction (instead of 94 in
the hydrodynamic models) and 128 in
$\phi$-direction (instead of 49 per quadrant in our hydrodynamic models) 
is currently feasible. We use 96 distinct wavelengths which are logarithmically 
distributed between 0.001 and 2000\,$\umu$m, with an enhanced resolution in the infrared regime
and especially around the silicate features. 
The gas temperature from the hydrodynamic simulation is only used
to decide whether dust is present at a certain grid point ($T_{\mathrm{gas}} < T_{sub}^{sil/graph}$) 
or not ($T_{\mathrm{gas}} \ge T_{sub}^{sil/graph}$). As we do not follow the dust temperature during 
the hydrodynamical simulation explicitly, we set the dust temperature to zero prior to the 
radiative transfer calculation and assume that the dust distribution is solely heated by the central AGN.
We assume a bolometric luminosity of the accretion disk of $L_{\mathrm{disk}}=1.2\cdot 10^{11}\,L_{\odot}$, 
with a $\cos\,\theta$ radiation characteristic, preferentially emitting perpendicular to the plane of 
the disk.

Fig.~\ref{fig:mc3d_hydro} shows an inclination angle study for images at
$12\,\umu$m of the final stage (see Fig.~\ref{fig:a007_tsli_999.eps}) of our standard
model. Shown is the dust re-emission and the central radiation of the almost face-on case
($i=8\degr$)\footnote{An inclination angle of $8\degr$ is used in order to investigate only lines of sight 
within the computational domain of the hydrodynamical simulations as well as within the torus of 
our previously investigated 2D 
radiative transfer simulations we compare to in Sect.\,\ref{sec:feat_hi}.}, 
$i=30\degr$, $i=60\degr$ and the edge-on case ($i=90\degr$) for
an azimuthal angle of $\phi=45\degr$. 
Even at an inclination angle of $60\degr$, the central source still possesses
the highest surface brightness 
at these $\phi$-angles, but it vanishes behind the dense
disk component at larger inclination angles ($90\degr$). 
The dense dust disk is the second brightest characteristic
for our models. As expected, it does not appear smooth, but a filamentary
structure is visible, as well as a slight nuclear spiral structure
(only visible in the high-resolution electronic version),
caused by the differential rotation of the disk. 
Due to the large optical depth at intermediate inclination angles, the torus
shows an asymmetric structure with respect to the midplane, as only the
upper, directly illuminated funnel wall is visible.
The same holds true for the $60\degr$ inclination. Here, as well
as in the edge-on case, {\it shadows} of dense -- in most cases radially
inwards moving -- filaments are visible. The anisotropic radiation source
(preferentially emitting along the axis) also
contributes to the vertically elongated bright regions and enhances the dark
lanes in the midplane.   

\begin{figure}
  \centering
  \resizebox{\hsize}{!}{\includegraphics[angle=0]{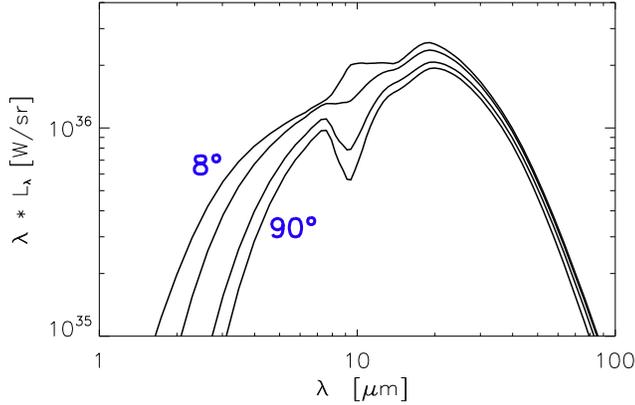}} 
  \caption[SEDs of our standard model]{SEDs
           of our standard model. Shown are the inclination angles
           $8\hbox{$^\circ$}$,  $30\hbox{$^\circ$}$,  $60\hbox{$^\circ$}$ and
           $90\hbox{$^\circ$}$ (top to bottom) for a common azimuthal angle of $\phi=45\hbox{$^\circ$}$.} 
  \label{fig:hydro_standard_SED.eps} 
\end{figure}

In Fig.~\ref{fig:hydro_standard_SED.eps}, SEDs of the standard
model are presented in terms of {\it pure} dust re-emission SEDs.
Due to the
turbulent structure of the dense disk component close to the midplane and the
filaments and blobs of high density gas within or close to the funnel region
(compare to Fig.~\ref{fig:a007_turbdisc_paper.eps}),
the model for the reduction of the silicate feature
in emission as described in \citet{Schartmann_08} works
fine and only a moderate emission band with a strength of 0.31 is
visible at $9.7\,\umu$m. For the following analysis, 
we define the silicate feature strength in accordance with
\citet{Shi_06} as 
\begin{eqnarray}
\label{equ:featchar}
\Delta_{\mathrm{feature}}=\frac{F_{\mathrm{feat}}-F_{\mathrm{cont}}}{F_{\mathrm{cont}}},
\end{eqnarray}
where $F_{\mathrm{feat}}$ is the flux at the wavelength of the silicate feature and 
$F_{\mathrm{cont}}$ is the spline fitted continuum, anchored at wavelengths between 5.0
and 7.5\,$\umu$m and from 25.0 to 40.0\,$\umu$m.
With increasing inclination angle, it changes to moderate absorption: -0.10 for $30\degr$, -0.41 for
$60\degr$ and -0.54 for $90\degr$.
This is remarkable, as the same average optical depth within the midplane
($<\tau_{9.7\,\umu\mathrm{m}}>_{\mathrm{equ}} \approx 400$!!)
would cause much deeper absorption -- which has never been observed --  
for the case of a torus with continuous
dust distribution. 
These small feature depths are caused by the fact that  
the silicate emission feature is
produced within an elongated volume along the funnel region. 
Thus, the appearance of the silicate feature is due to a mixture of emission and
absorption and, accordingly,
existing channels of low absorption towards the observer (though leading to
silicate emission features) can also contribute
to the final morphology of the silicate feature.

For a detailed discussion of the accuracy and limitations of these kind of radiative 
transfer calculations, 
we refer the reader to \citet{Schartmann_08}. The most severe problems 
occur in the midplane, where the largest
optical depths are reached, which lead to an overestimation of the temperature due to
steep gradients. Here, a higher spatial resolution would be desirable. However, 
we are confident that our generic results are robust, since a resolution
study for the case of SEDs yielded almost identical results, 
as also described in \citet{Schartmann_08}.


\subsection{Comparison with {\it Spitzer} spectra}

\begin{figure}
  \centering
  \resizebox{\linewidth}{!}{\includegraphics{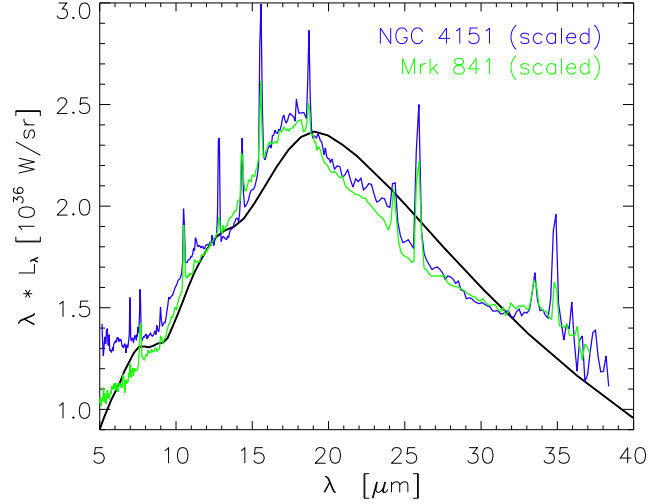}} 
 \caption[Comparison of our standard model with IRS {\it Spitzer}
  observations]{Comparison 
  of our standard model (solid line -- $i=30\degr$) with IRS {\it Spitzer} observations of
           NGC\,4151 \citep{Weedman_05} and Mrk\,841 (H.~W.~W.~Spoon, private communication).} 
  \label{fig:spitzer_comp.eps}
\end{figure}

Figure~\ref{fig:spitzer_comp.eps} shows a comparison of our standard
hydrodynamic model with observations of two Seyfert galaxies with the   
Infra-Red Spectrograph (IRS) onboard the {\it Spitzer Space Telescope}, 
in high-sensitivity, low resolution
mode. The solid curve refers to an inclination angle of $30\degr$ 
towards our standard model.   
Shown in colour are the mid-infrared SEDs of 
the two intermediate type (Sy\,1.5 according to
NED\footnote{http://nedwww.ipac.caltech.edu}, the NASA/IPAC Extragalactic
Database) Seyfert galaxies NGC\,4151 (blue solid line, \citealp{Weedman_05}) and
Mrk\,841 (green, H.~W.~W.~Spoon, private communication).
Both are scaled, in order to obtain 
fluxes similar to our standard model, as we are only interested in a comparison of
the shape of the curves. 
The $30\degr$-inclination angle of our simulated SED shows a self-absorbed
silicate feature, yielding a good approximation of the green curve in the 
respective wavelength range.
As we are applying pure dust radiation transfer, we are not able
to produce emission lines in our modelled SEDs and we can only compare
the underlying continuum flux. 
Deviations between observations and model are
visible in a shift of the 
maximum of emission of our model towards slightly longer wavelengths and a
feature around $34\,\umu$m. The latter arises most likely from crystalline 
forsterite in emission (H.~W.~W.~Spoon, private communication), which is not 
included in our dust model. 

As our SEDs do not result from a fitting procedure, this is a very noteworthy
result. Especially because the physical parameters we use are meant to represent
a typical Seyfert galaxy. The sharp decrease of the flux towards longer wavelengths results from
missing cold dust in the outer part.


\subsection{Silicate feature strength H{\sc \,i} column density relation}
\label{sec:feat_hi}

Compared to the previously discussed infrared wavelength range, the opacity 
drops steeply in the X-ray range. 
Therefore, 
a certain amount of X-ray photons can escape and provides thus the
possibility to
determine the hydrogen column density on the line of sight observationally.

For the case of our hydrodynamic models, H{\sc \,i} column
densities are directly integrated along the line of sight. Only cells are
considered, where the temperature is lower than the ionisation temperature of
hydrogen.  
However, in our continuous models to which we will compare our new hydrodynamical models
in the following, we cannot take this into account. Therefore, we assume a
constant gas-to-dust ratio of 250 (see discussion in
Sect.~\ref{sec:hydro_observable}), 
in order to calculate column densities from
the continuous dust distributions. 
The strength of the silicate feature is derived as given 
in formula \ref{equ:featchar} and
is taken at the wavelength, where the maximum distance between the
SED and the fitted continuum is reached, as it can be appreciably shifted
for the case of self-absorbed emission. 
\begin{figure}
  \centering
  \resizebox{\linewidth}{!}{\includegraphics{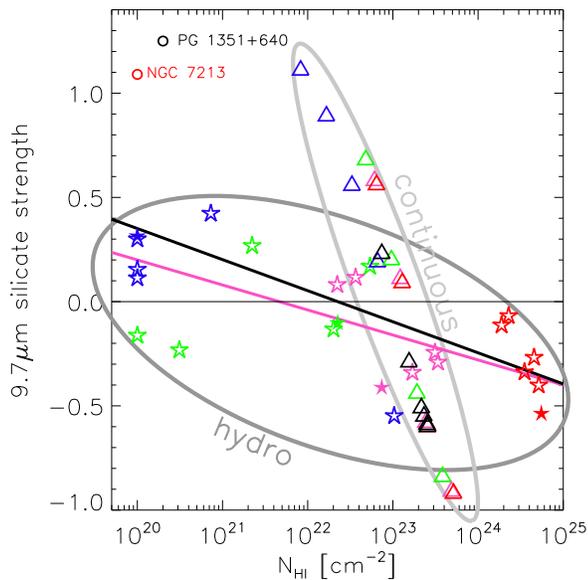}} 
  \caption[Silicate feature strengths and column
           densities of our hydromodels]{Comparison of resulting silicate feature strengths and column
           densities of our hydromodels (standard model plus models from a supernova and mass loss rate study)
           given as stars and the continuous TTM-models (standard model
           and models from a dust mass study, see \citealp{Schartmann_05}) given as
           triangles. The filled stars denote our hydrodynamical standard
           model. Colour denotes inclination angle (blue -- $8\degr$, green
           -- $30\degr$, magenta -- $60\degr$, red -- $90\degr$, black -- intermediate angles of the
           continuous standard model). 
           The solid lines
           are linear fits to the observational data shown in
           \citealp{Shi_06} (black -- all objects, magenta -- Seyfert
           galaxy sample). The two outlying objects NGC\,7213 and PG\,1351+640
           are given additionally. Ellipses are drawn to guide the eye.} 
  \label{fig:shi_comparison.eps} 
\end{figure}
With the help of this procedure, our simulations can be directly compared 
to the observational data published in \citet{Shi_06}. 
The \citet{Shi_06} sample
comprises of 85 AGN of various types, for which {\it Spitzer} InfraRed
Spectrograph \citep[IRS,][]{Houck_04} observations or 
ground-based measurements from the \citet{Roche_91} sample (four objects)
exist, as well as X-ray data. 
H{\sc \,i} column densities are obtained from X-ray spectra taken from the
Chandra data archive directly for 8 objects. For further 77 objects, column
densities were collected from the literature. 
A correlation was
found between the silicate feature strength (see equation
\ref{equ:featchar}) and the X-ray absorption column density in a sense
that low H{\sc \,i} columns correspond to silicate emission and high columns
correspond to silicate absorption:  
\begin{eqnarray}
\label{equ:featstrength_data}
\Delta_{\mathrm{feature}}=3.3\pm0.5-(0.15\pm0.02)\cdot \log(N_{\mathrm{HI}}).
\end{eqnarray}
This is shown as the black line in Fig.~\ref{fig:shi_comparison.eps}. 
When \citet{Shi_06} determine the correlation between the silicate feature strength and
the X-ray column density data separately for 
the various types of AGN, a large scatter between them is found,
as is also visible in the data range for each type
of AGN. 
Indicated in magenta is the correlation for Seyfert galaxies alone, given by:
\begin{eqnarray}
\label{equ:featstrength_data_seyf}
\Delta_{\mathrm{feature}}=2.6\pm0.7-(0.12\pm0.03)\cdot \log(N_{\mathrm{HI}}).
\end{eqnarray}
 
In Fig.~\ref{fig:shi_comparison.eps}, these correlations are compared to our 
simulations. 
Colours of the symbols indicate inclination angle:
blue -- $8\degr$, green -- $30\degr$, magenta -- $60\degr$ and red -- $90\degr$.
For the case of inclination angles without any dust along the 
line-of-sight, we assigned a value of $N_{\mathrm{HI}}=10^{20}\,cm^{-2}$,
taking foreground gas into account. 
Our hydrodynamic models are given as stars and the triangles 
denote results from our continuous TTM-model simulations \citep{Schartmann_05}.
Overlayed ellipses roughly mark the distribution of symbols for the two 
classes of simulations.  
Excluding the values at $10^{20}\,cm^{-2}$, which are only due to foreground extinction,
one can see that the hydrodynamical models agree quite well with the observed
correlations, whereas the continuous models show a much too steep dependence.
Although the emission features in the Seyfert\,1 case are rather too
pronounced, they are in good agreement with the two outlying galaxies NGC\,7213
and PG\,1351+640. Furthermore, the silicate feature changes from emission to absorption
within a very small range of column densities and it is also evident that column
densities much higher than $10^{24}\,cm^{-2}$ cannot be reached, as they would
cause unphysically high absorption depths of the silicate feature
(as already discussed in \citealp{Schartmann_08}). A further
concentration of the dust density towards the minimum of the potential within
the midplane might cause a flattening of this distribution, but then, too
narrow infrared bumps will be produced, see \citet{Schartmann_03}. 
Obviously, the observations presented in \citet{Shi_06} are in favour of
clumpy or filamentary models instead of continuous matter distributions, 
concerning the models we investigated so
far. At least, models are needed which feature high column densities in
combination with moderate emission as well as absorption features. In the
hydrodynamical models presented in this paper, this is possible due to the
clumpy and filamentary nature of the dust distribution, which enables the
coexistence of almost free lines of sight towards the inner region, where the
silicate feature in emission is produced, together with 
highly obscured lines of sight. The highest column densities are reached for
lines of sight through the dense and turbulent disk in our models.
Furthermore, a remarkable consequence of this is
the very tight linear correlation of the continuous models and the broad
distribution for the case of our (clumpy) hydrodynamical models.
Therefore, the broad scatter in the observational data
might really be taken as a case for clumpy models, as claimed in
\citet{Shi_06}. 
The observed shallow slope of the discussed relation has also
been investigated by \citet{Levenson_07} on basis of 
radiative transfer models of simplified geometries, representing dust
clouds and continuous dust distributions. Whereas the latter can yield
arbitrarily deep silicate absorption features in their simulations, 
single dust clouds can not, as long as their dimensions are much smaller than
their distance to the illuminating source. With the help of these simulations
they conclude that the tori of Seyfert galaxies must be clumpy, as their mean
SED shows only moderate emission as well as absorption features
and ULIRGs ({\it Ultraluminous Infrared Galaxies}) must possess nuclear sources, which are deeply embedded
into a geometrically and optically thick and smooth distribution of dust,
as their mean SED features deep silicate absorption \citep{Hao_07}.
Similar conclusions have been drawn with the help of a clumpy torus model
already discussed in section \ref{sec:intro} by \cite{Nenkova_02}
and \citet{Nenkova_08a,Nenkova_08b}.

\section{Summary}
\label{sec:summary}

 We present a physical model for the evolution of gaseous tori in Seyfert galaxies,
 in which the stellar evolution of a young nuclear star cluster
is predominantly responsible for setting up a torus-like
dust (and gas) distribution. The star cluster provides both, 
(i) mass injection, which is mainly due to planetary nebulae and (ii) injection
of energy from supernova explosions.
 Within the course of the simulations, a highly dynamical 
 multiphase medium evolves, which is comparable to
 high resolution models of the supernova-driven turbulent ISM in galaxies 
 \citep{Breitschwerdt_07,Joung_06}. 
 There, blast waves from supernovae explosions sweep up
 the ISM by transferring kinetic energy to the gas, which leads to the
 subsequent formation of thin shells. 
 These shells interact with one another and with already existing 
 filaments. Radiative cooling then produces cold clumps. 
 In our scenario, the ISM
 is constantly refilled by stellar mass loss processes 
 distributed according to the 
stars in the nuclear star cluster. The
 main structuring agent of the gas and dust in the torus is the interplay between mass input, 
 supernova blast waves locally sweeping away the ISM, the streams of 
 cold gas towards the centre as well as cooling instability. \\
 Within this scenario, cold dense clouds and filaments, 
 where dust can exist, form naturally. 
 The implemented optically thin cooling reduces thermal pressure and, therefore,
 material streams towards the minimum of
 the potential. This forms long radial filaments, with lots of substructure
 due to interaction processes.   
 In agreement with MHD simulations of the ISM by \citet{Breitschwerdt_07}, 
 we find absence of pressure equilibrium of the temperature phases. 
 This is the case, as mixing due to supersonic motions and turbulence as well
 as cooling processes act on shorter time-scales than relaxation processes.
 Although no pressure equilibrium can be reached, the flow
 is in a global dynamical equilibrium state. The
 characteristics of this equilibrium delicately depend on the driving
 parameters -- that are energy input (supernova rate) and mass input rate, as well as
 the gravitational potential, the cooling curve and
 rotation. 
 We find completely different velocities and streaming patterns
 for the various temperature phases. Whereas cold material generally 
 tends to move towards the
 centre, the radial velocity of the hot ionised medium shows both inflow (in 
 the inner region) and outflow (further out). 
 Depending on the ratio between supernova rate and mass loss rate, the
inflow/outflow transition shifts towards smaller radii for higher 
SN-to-mass injection ratios and towards larger
 radii, 
 when mass input dominates over supernova heating. 
 For very high supernova rates, we observe solely outflow solutions for
 the hot gas component. This solution is likely to dominate 
 the initial supernova type~II phase
 shortly after the birth of the stellar population. 

 The resulting density and temperature structure of the gas is subsequently
 fed into the radiative transfer code {\sc MC3D} in order to obtain observable
 quantities (SEDs and surface brightness distributions in the mid-infrared), 
 which are then compared to recent observations. 
 A comparison with spectral energy data from the {\it Spitzer} satellite shows that,
despite of its severe limitations in resolution and astrophysics considered
within this exploratory study,
 our hydrodynamical standard model
 describes one class of {\it Spitzer} SEDs fairly
 well, without the necessity of any finetuning of
 parameters. For example the two intermediate type Seyfert galaxies NGC\,4151 
 as well as Mrk~841 belong to this
 class (see Fig.~\ref{fig:spitzer_comp.eps}).  
 The success of the comparison with the silicate feature strength to H{\sc i}
 column density relation
 for the hydrodynamical models and the failure of the continuous
 TTM-models leads to the important conclusion that the presence of a 
 disk-like density
 enhancement within the equatorial plane in combination with a fluffy 
 structure surrounding it is needed in order to get agreement
 with observational data. Otherwise, large X-ray column densities cannot
 simultaneously be obtained together with moderate  
 silicate absorption feature strengths at viewing
 angles within the midplane. The latter are guaranteed by a mixture of almost dust-free
 and heavily obscured lines of sight towards the hot inner part of the torus close to the funnel.
 Therefore, we see a much weaker silicate absorption feature. 
 Due to these remarkable similarities of our exploratory study and observations, we feel 
encouraged to implement more physical effects into our simulations and to study
the central gas and dust structure over a longer evolutionary time of the nuclear star cluster.

\section*{Acknowledgments}

We would like to thank the anonymous referee for many useful suggestions to improve the paper, 
H.W.W.\,Spoon for providing us the reduced {\it Spitzer} SED of Mrk~841
prior to publication, as well as K.R.W.\,Tristram and M.\,Krause for helpful discussion.
S.\,W.\,was supported by the German 
Research Foundation (DFG) through the Emmy Noether grant WO\,857/2.

\bibliographystyle{mn2e}
\bibliography{astrings,literature}

\begin{thebibliography}{}

\bibitem[\protect\citeauthoryear{{Antonucci}}{{Antonucci}}{1993}]{Antonucci_93}
{Antonucci} R.,  1993, Ann. Rev. Astron. Astrophys., 31, 473

\bibitem[\protect\citeauthoryear{{Antonucci} \& {Miller}}{{Antonucci} \&
  {Miller}}{1985}]{Antonucci_85}
{Antonucci} R.~R.~J.,  {Miller} J.~S.,  1985, Astrophys. J., 297, 621

\bibitem[\protect\citeauthoryear{{Balbus} \& {Hawley}}{{Balbus} \&
  {Hawley}}{1991}]{Balbus_91}
{Balbus} S.~A.,  {Hawley} J.~F.,  1991, Astrophys. J., 376, 214

\bibitem[\protect\citeauthoryear{{Beckert} \& {Duschl}}{{Beckert} \&
  {Duschl}}{2004}]{Beckert_04}
{Beckert} T.,  {Duschl} W.~J.,  2004, Astron. Astrophys., 426, 445

\bibitem[\protect\citeauthoryear{{Bender}, {Kormendy}, {Bower}, {Green},
  {Thomas}, {Danks}, {Gull}, {Hutchings}, {Joseph}, {Kaiser}, {Lauer},
  {Nelson}, {Richstone}, {Weistrop} \& {Woodgate}}{{Bender}
  et~al.}{2005}]{Bender_05}
{Bender} R.,  {Kormendy} J.,  {Bower} G.,  {Green} R.,  {Thomas} J.,  {Danks}
  A.~C.,  {Gull} T.,  {Hutchings} J.~B.,  {Joseph} C.~L.,  {Kaiser} M.~E.,
  {Lauer} T.~R.,  {Nelson} C.~H.,  {Richstone} D.,  {Weistrop} D.,
  {Woodgate} B.,  2005, Astrophys. J., 631, 280

\bibitem[\protect\citeauthoryear{{Breitschwerdt} \& {de
  Avillez}}{{Breitschwerdt} \& {de Avillez}}{2007}]{Breitschwerdt_07}
{Breitschwerdt} D.,  {de Avillez} M.~A.,  2007, in {Elmegreen} B.~G.,  {Palous}
  J.,  eds, IAU Symposium Vol.~237 of IAU Symposium, {Dynamical evolution of a
  supernova driven turbulent interstellar medium}.
pp 57--64

\bibitem[\protect\citeauthoryear{{Bressan}, {Fagotto}, {Bertelli} \&
  {Chiosi}}{{Bressan} et~al.}{1993}]{Bressan_93}
{Bressan} A.,  {Fagotto} F.,  {Bertelli} G.,    {Chiosi} C.,  1993, Astron.
  Astrophys. Suppl. Ser., 100, 647

\bibitem[\protect\citeauthoryear{{Camenzind}}{{Camenzind}}{1995}]{Camenzind_95}
{Camenzind} M.,  1995, Reviews of Modern Astronomy, 8, 201

\bibitem[\protect\citeauthoryear{{Courant} \& {Friedrichs}}{{Courant} \&
  {Friedrichs}}{1948}]{Courant_48}
{Courant} R.,  {Friedrichs} K.~O.,  1948, {Supersonic flow and shock waves}.
Pure and Applied Mathematics, New York: Interscience, 1948

\bibitem[\protect\citeauthoryear{{Dalgarno} \& {McCray}}{{Dalgarno} \&
  {McCray}}{1972}]{Dalgarno_72}
{Dalgarno} A.,  {McCray} R.~A.,  1972, Ann. Rev. Astron. Astrophys., 10, 375

\bibitem[\protect\citeauthoryear{{Davies}, {Mueller S{\'a}nchez}, {Genzel},
  {Tacconi}, {Hicks}, {Friedrich} \& {Sternberg}}{{Davies}
  et~al.}{2007}]{Davies_07}
{Davies} R.~I.,  {Mueller S{\'a}nchez} F.,  {Genzel} R.,  {Tacconi} L.~J.,
  {Hicks} E.~K.~S.,  {Friedrich} S.,    {Sternberg} A.,  2007, Astrophys. J.,
  671, 1388

\bibitem[\protect\citeauthoryear{{de Avillez} \& {Breitschwerdt}}{{de Avillez}
  \& {Breitschwerdt}}{2004}]{Avillez_04}
{de Avillez} M.,  {Breitschwerdt} D.,  2004, Astrophys. Space. Sci., 292, 207

\bibitem[\protect\citeauthoryear{{de Avillez} \& {Breitschwerdt}}{{de Avillez}
  \& {Breitschwerdt}}{2005}]{Avillez_05}
{de Avillez} M.~A.,  {Breitschwerdt} D.,  2005, Astron. Astrophys., 436, 585

\bibitem[\protect\citeauthoryear{{de Donder} \& {Vanbeveren}}{{de Donder} \&
  {Vanbeveren}}{2003}]{DeDonder_03}
{de Donder} E.,  {Vanbeveren} D.,  2003, New Astronomy, 8, 817

\bibitem[\protect\citeauthoryear{{Draine} \& {Lee}}{{Draine} \&
  {Lee}}{1984}]{Draine_84}
{Draine} B.~T.,  {Lee} H.~M.,  1984, Astrophys. J., 285, 89

\bibitem[\protect\citeauthoryear{{Elitzur}}{{Elitzur}}{2006}]{Elitzur_06}
{Elitzur} M.,  2006, New Astronomy Review, 50, 728

\bibitem[\protect\citeauthoryear{{Elmegreen}}{{Elmegreen}}{2005}]{Elmegreen_05}
{Elmegreen} B.~G.,  2005, in {de Grijs} R.,  {Gonz{\'a}lez Delgado} R.~M.,
  eds, ASSL Vol. 329: Starbursts: From 30 Doradus to Lyman Break Galaxies {The
  Initial Mass Function in Starbursts}.
p.~57

\bibitem[\protect\citeauthoryear{{Ferland}}{{Ferland}}{1993}]{Ferland_93}
{Ferland} G.~J.,  1993, {Hazy, A Brief Introduction to Cloudy 84}.
University of Kentucky Internal Report, 565 pages

\bibitem[\protect\citeauthoryear{{Gaibler}, {Camenzind} \& {Krause}}{{Gaibler}
  et~al.}{2005}]{Gaibler_05}
{Gaibler} V.,  {Camenzind} M.,    {Krause} M.,  2005, in {Merloni} A.,
  {Nayakshin} S.,   {Sunyaev} R.~A.,  eds, Growing Black Holes: Accretion in a
  Cosmological Context {Evolution of the ISM in elliptical galaxies and black
  hole growth}.
pp 66--67

\bibitem[\protect\citeauthoryear{{Gallimore} \& {Matthews}}{{Gallimore} \&
  {Matthews}}{2003}]{Gallimore_03}
{Gallimore} J.~F.,  {Matthews} L.,  2003, in ASP Conf. Ser. 290: Active
  Galactic Nuclei: From Central Engine to Host Galaxy {NICMOS Observations of
  the Nuclear Star Cluster of NGC 1068}.
p.~501

\bibitem[\protect\citeauthoryear{{Gerola}, {Kafatos} \& {McCray}}{{Gerola}
  et~al.}{1974}]{Gerola_74}
{Gerola} H.,  {Kafatos} M.,    {McCray} R.,  1974, Astrophys. J., 189, 55

\bibitem[\protect\citeauthoryear{{Greenhill}, {Booth}, {Ellingsen},
  {Herrnstein}, {Jauncey}, {McCulloch}, {Moran}, {Norris}, {Reynolds} \&
  {Tzioumis}}{{Greenhill} et~al.}{2003}]{Greenhill_03}
{Greenhill} L.~J.,  {Booth} R.~S.,  {Ellingsen} S.~P.,  {Herrnstein} J.~R.,
  {Jauncey} D.~L.,  {McCulloch} P.~M.,  {Moran} J.~M.,  {Norris} R.~P.,
  {Reynolds} J.~E.,    {Tzioumis} A.~K.,  2003, Astrophys. J., 590, 162

\bibitem[\protect\citeauthoryear{{Hao}, {Weedman}, {Spoon}, {Marshall},
  {Levenson}, {Elitzur} \& {Houck}}{{Hao} et~al.}{2007}]{Hao_07}
{Hao} L.,  {Weedman} D.~W.,  {Spoon} H.~W.~W.,  {Marshall} J.~A.,  {Levenson}
  N.~A.,  {Elitzur} M.,    {Houck} J.~R.,  2007, Astrophys. J., Lett., 655, L77

\bibitem[\protect\citeauthoryear{{H{\"o}nig}, {Beckert}, {Ohnaka} \&
  {Weigelt}}{{H{\"o}nig} et~al.}{2006}]{Hoenig_06}
{H{\"o}nig} S.~F.,  {Beckert} T.,  {Ohnaka} K.,    {Weigelt} G.,  2006, Astron.
  Astrophys., 452, 459

\bibitem[\protect\citeauthoryear{{Houck}, {Roellig}, {van Cleve}, {Forrest},
  {Herter}, {Lawrence}, {Matthews}, {Reitsema}, {Soifer}, {Watson}, {Weedman},
  {Huisjen}, {Troeltzsch}, {Barry}, {Bernard-Salas}, {Blacken}, {Brandl},
  {Charmandaris}, {Devost}, {Gull}, {Hall},  {Henderson},  
  {Higdon}, {Pirger}, {Schoenwald}, {Sloan},
  {Uchida}, {Appleton}, {Armus}, {Burgdorf},
  {Fajardo-Acosta}, {Grillmair}, {Ingalls}, {Morris}, {Teplitz}}{Houck et~al.}{2004}]{Houck_04}
{Houck} J.~R.,  {Roellig} T.~L.,  {van Cleve} J.,  {Forrest} W.~J.,  {Herter}
  T.,  {Lawrence} C.~R.,  {Matthews} K.,  {Reitsema} H.~J.,  {Soifer} B.~T.,
  {Watson} D.~M.,  {Weedman} D.,  {Huisjen} M.,  {Troeltzsch} J.,  {Barry}
  D.~J.,  {Bernard-Salas} J.,  {Blacken} C.~E.,  {Brandl} B.~R.,
  {Charmandaris} V.,  {Devost} D.,  {Gull} G.~E.,  {Hall} P.,  {Henderson}
  C.~P.,  {Higdon} S.~J.~U.,  {Pirger} B.~E.,  {Schoenwald} J.,  {Sloan} G.~C.,
   {Uchida} K.~I.,  {Appleton} P.~N.,  {Armus} L.,  {Burgdorf} M.~J.,
  {Fajardo-Acosta} S.~B.,  {Grillmair} C.~J.,  {Ingalls} J.~G.,  {Morris}
  P.~W.,    {Teplitz} H.~I.,  2004, Astrophys. J., Suppl. Ser., 154, 18

\bibitem[\protect\citeauthoryear{{Jaffe}, {Meisenheimer}, {R{\" o}ttgering},
  {Leinert}, {Richichi}, {Chesneau}, {Fraix-Burnet}, {Glazenborg-Kluttig},
  {Granato}, {Graser}, {Heijligers}, {K{\" o}hler}, {Malbet}, {Miley},
  {Paresce}, {Pel}, {Perrin}, {Przygodda}, {Schoeller}, {Sol}, {Waters},
  {Weigelt}, {Woillez} \& {de Zeeuw}}{{Jaffe} et~al.}{2004}]{Jaffe_04}
{Jaffe} W.,  {Meisenheimer} K.,  {R{\" o}ttgering} H.~J.~A.,  {Leinert} C.,
  {Richichi} A.,  {Chesneau} O.,  {Fraix-Burnet} D.,  {Glazenborg-Kluttig} A.,
  {Granato} G.-L.,  {Graser} U.,  {Heijligers} B.,  {K{\" o}hler} R.,  {Malbet}
  F.,  {Miley} G.~K.,  {Paresce} F.,  {Pel} J.-W.,  {Perrin} G.,  {Przygodda}
  F.,  {Schoeller} M.,  {Sol} H.,  {Waters} L.~B.~F.~M.,  {Weigelt} G.,
  {Woillez} J.,    {de Zeeuw} P.~T.,  2004, Nature, 429, 47

\bibitem[\protect\citeauthoryear{{Jenkins} \& {Tripp}}{{Jenkins} \&
  {Tripp}}{2007}]{Jenkins_07}
{Jenkins} E.~B.,  {Tripp} T.~M.,  2007, in {Elmegreen} B.~G.,  {Palous} J.,
  eds, IAU Symposium Vol.~237 of IAU Symposium, {New results on the
  distribution of thermal pressures in the diffuse ISM}.
pp 53--56

\bibitem[\protect\citeauthoryear{{Joung} \& {Mac Low}}{{Joung} \& {Mac
  Low}}{2006}]{Joung_06}
{Joung} M.~K.~R.,  {Mac Low} M.-M.,  2006, Astrophys. J., 653, 1266

\bibitem[\protect\citeauthoryear{{Jungwiert}, {Combes} \& {Palou{\v
  s}}}{{Jungwiert} et~al.}{2001}]{Jungwiert_01}
{Jungwiert} B.,  {Combes} F.,    {Palou{\v s}} J.,  2001, Astron. Astrophys.,
  376, 85

\bibitem[\protect\citeauthoryear{{Klahr}}{{Klahr}}{1998}]{Klahr_98}
{Klahr} H.~H.,  1998, PhD thesis, Friedrich-Schiller-Universit\"at Jena

\bibitem[\protect\citeauthoryear{{Klahr}, {Henning} \& {Kley}}{{Klahr}
  et~al.}{1999}]{Klahr_99}
{Klahr} H.~H.,  {Henning} T.,    {Kley} W.,  1999, Astrophys. J., 514, 325

\bibitem[\protect\citeauthoryear{{Kley}}{{Kley}}{1989}]{Kley_89}
{Kley} W.,  1989, Astron. Astrophys., 208, 98

\bibitem[\protect\citeauthoryear{{K\"onigl} \& {Kartje}}{{K\"onigl} \&
  {Kartje}}{1994}]{Koenigl_94}
{K\"onigl} A.,  {Kartje} J.~F.,  1994, Astrophys. J., 434, 446

\bibitem[\protect\citeauthoryear{{Krolik}}{{Krolik}}{2007}]{Krolik_07}
{Krolik} J.~H.,  2007, Astrophys. J., 661, 52

\bibitem[\protect\citeauthoryear{{Krolik} \& {Begelman}}{{Krolik} \&
  {Begelman}}{1988}]{Krolik_88}
{Krolik} J.~H.,  {Begelman} M.~C.,  1988, Astrophys. J., 329, 702

\bibitem[\protect\citeauthoryear{{Kwok}}{{Kwok}}{2005}]{Kwok_05}
{Kwok} S.,  2005, Journal of Korean Astronomical Society, 38, 271

\bibitem[\protect\citeauthoryear{{Laor} \& {Draine}}{{Laor} \&
  {Draine}}{1993}]{Laor_93}
{Laor} A.,  {Draine} B.~T.,  1993, Astrophys. J., 402, 441

\bibitem[\protect\citeauthoryear{{Leitherer}, {Schaerer}, {Goldader},
  {Delgado}, {Robert}, {Kune}, {de Mello}, {Devost} \& {Heckman}}{{Leitherer}
  et~al.}{1999}]{Leitherer_99}
{Leitherer} C.,  {Schaerer} D.,  {Goldader} J.~D.,  {Delgado} R.~M.~G.,
  {Robert} C.,  {Kune} D.~F.,  {de Mello} D.~F.,  {Devost} D.,    {Heckman}
  T.~M.,  1999, Astrophys. J., Suppl. Ser., 123, 3

\bibitem[\protect\citeauthoryear{{Levenson}, {Sirocky}, {Hao}, {Spoon},
  {Marshall}, {Elitzur} \& {Houck}}{{Levenson} et~al.}{2007}]{Levenson_07}
{Levenson} N.~A.,  {Sirocky} M.~M.,  {Hao} L.,  {Spoon} H.~W.~W.,  {Marshall}
  J.~A.,  {Elitzur} M.,    {Houck} J.~R.,  2007, Astrophys. J., Lett., 654, L45

\bibitem[\protect\citeauthoryear{{Levermore} \& {Pomraning}}{{Levermore} \&
  {Pomraning}}{1981}]{Levermore_81}
{Levermore} C.~D.,  {Pomraning} G.~C.,  1981, Astrophys. J., 248, 321

\bibitem[\protect\citeauthoryear{{Maciejewski}}{{Maciejewski}}{2006}]{Maciejew%
ski_06}
{Maciejewski} W.,  2006, astro-ph/0611259

\bibitem[\protect\citeauthoryear{{Marigo}, {Bressan} \& {Chiosi}}{{Marigo}
  et~al.}{1996}]{Marigo_96}
{Marigo} P.,  {Bressan} A.,    {Chiosi} C.,  1996, Astron. Astrophys., 313, 545

\bibitem[\protect\citeauthoryear{{Mathis}, {Rumpl} \& {Nordsieck}}{{Mathis}
  et~al.}{1977}]{Mathis_77}
{Mathis} J.~S.,  {Rumpl} W.,    {Nordsieck} K.~H.,  1977, Astrophys. J., 217,
  425

\bibitem[\protect\citeauthoryear{{Mignone}, {Bodo}, {Massaglia}, {Matsakos},
  {Tesileanu}, {Zanni} \& {Ferrari}}{{Mignone} et~al.}{2007}]{Mignone_07}
{Mignone} A.,  {Bodo} G.,  {Massaglia} S.,  {Matsakos} T.,  {Tesileanu} O.,
  {Zanni} C.,    {Ferrari} A.,  2007, Astrophys. J., Suppl. Ser., 170, 228

\bibitem[\protect\citeauthoryear{{Nayakshin}, {Dehnen}, {Cuadra} \&
  {Genzel}}{{Nayakshin} et~al.}{2006}]{Nayakshin_06a}
{Nayakshin} S.,  {Dehnen} W.,  {Cuadra} J.,    {Genzel} R.,  2006, Mon. Not. R.
  Astron. Soc., 366, 1410

\bibitem[\protect\citeauthoryear{{Nenkova}, {Ivezi{\' c}} \&
  {Elitzur}}{{Nenkova} et~al.}{2002}]{Nenkova_02}
{Nenkova} M.,  {Ivezi{\' c}} {\v Z}.,    {Elitzur} M.,  2002, Astrophys. J.,
  Lett., 570, L9

\bibitem[\protect\citeauthoryear{{Nenkova}, {Sirocky}, {Ivezic} \&
  {Elitzur}}{{Nenkova} et~al.}{2008a}]{Nenkova_08a}
{Nenkova} M.,  {Sirocky} M.~M.,  {Ivezic} Z.,    {Elitzur} M.,  2008a, ArXiv
  e-prints, 806

\bibitem[\protect\citeauthoryear{{Nenkova}, {Sirocky}, {Nikutta}, {Ivezic} \&
  {Elitzur}}{{Nenkova} et~al.}{2008b}]{Nenkova_08b}
{Nenkova} M.,  {Sirocky} M.~M.,  {Nikutta} R.,  {Ivezic} Z.,    {Elitzur} M.,
  2008b, ArXiv e-prints, 806

\bibitem[\protect\citeauthoryear{{Paumard}, {Genzel}, {Martins}, {Nayakshin},
  {Beloborodov}, {Levin}, {Trippe}, {Eisenhauer}, {Ott}, {Gillessen}, {Abuter},
  {Cuadra}, {Alexander} \& {Sternberg}}{{Paumard} et~al.}{2006}]{Paumard_06}
{Paumard} T.,  {Genzel} R.,  {Martins} F.,  {Nayakshin} S.,  {Beloborodov}
  A.~M.,  {Levin} Y.,  {Trippe} S.,  {Eisenhauer} F.,  {Ott} T.,  {Gillessen}
  S.,  {Abuter} R.,  {Cuadra} J.,  {Alexander} T.,    {Sternberg} A.,  2006,
  Astrophys. J., 643, 1011

\bibitem[\protect\citeauthoryear{{Pier} \& {Krolik}}{{Pier} \&
  {Krolik}}{1992}]{Pier_92}
{Pier} E.~A.,  {Krolik} J.~H.,  1992, Astrophys. J., Lett., 399, L23

\bibitem[\protect\citeauthoryear{{Plewa}}{{Plewa}}{1995}]{Plewa_95}
{Plewa} T.,  1995, Mon. Not. R. Astron. Soc., 275, 143

\bibitem[\protect\citeauthoryear{{Plummer}}{{Plummer}}{1911}]{Plummer_11}
{Plummer} H.~C.,  1911, Mon. Not. R. Astron. Soc., 71, 460

\bibitem[\protect\citeauthoryear{{Poncelet}, {Perrin} \& {Sol}}{{Poncelet}
  et~al.}{2006}]{Poncelet_06}
{Poncelet} A.,  {Perrin} G.,    {Sol} H.,  2006, Astron. Astrophys., 450, 483

\bibitem[\protect\citeauthoryear{{Prieto}, {Maciejewski} \&
  {Reunanen}}{{Prieto} et~al.}{2005}]{Prieto_05}
{Prieto} M.~A.,  {Maciejewski} W.,    {Reunanen} J.,  2005, The Astron. J.,
  130, 1472

\bibitem[\protect\citeauthoryear{{Raban}, {Jaffe}, {R\"ottgering},
  {Meisenheimer} \& {Tristram}}{{Raban} et~al.}{2008}]{Raban_08}
{Raban} D.,  {Jaffe} W.,  {R\"ottgering} H.,  {Meisenheimer} K.,    {Tristram}
  K.~R.~W.,  2008, Astron. Astrophys.

\bibitem[\protect\citeauthoryear{{Risaliti}, {Elvis} \& {Nicastro}}{{Risaliti}
  et~al.}{2002}]{Risaliti_02}
{Risaliti} G.,  {Elvis} M.,    {Nicastro} F.,  2002, Astrophys. J., 571, 234

\bibitem[\protect\citeauthoryear{{Ritchmyer} \& {Morton}}{{Ritchmyer} \&
  {Morton}}{1967}]{Ritchmyer_67}
{Ritchmyer} R.~D.,  {Morton} K.~W.,  1967, {Difference methods for
  initial-value problems}.
Interscience Tracts in Pure and Applied Mathematics, New York: Interscience,
  1967, 2nd ed.

\bibitem[\protect\citeauthoryear{{Roche}, {Aitken}, {Smith} \& {Ward}}{{Roche}
  et~al.}{1991}]{Roche_91}
{Roche} P.~F.,  {Aitken} D.~K.,  {Smith} C.~H.,    {Ward} M.~J.,  1991, Mon.
  Not. R. Astron. Soc., 248, 606

\bibitem[\protect\citeauthoryear{{Salpeter}}{{Salpeter}}{1955}]{Salpeter_55}
{Salpeter} E.~E.,  1955, Astrophys. J., 121, 161

\bibitem[\protect\citeauthoryear{{Sanders}, {Phinney}, {Neugebauer}, {Soifer}
  \& {Matthews}}{{Sanders} et~al.}{1989}]{Sanders_89}
{Sanders} D.~B.,  {Phinney} E.~S.,  {Neugebauer} G.,  {Soifer} B.~T.,
  {Matthews} K.,  1989, Astrophys. J., 347, 29

\bibitem[\protect\citeauthoryear{{Scalo}}{{Scalo}}{1990}]{Scalo_90}
{Scalo} J.,  1990, in {Fabbiano} G.,  {Gallagher} J.~S.,   {Renzini} A.,  eds,
  ASSL Vol. 160: Windows on Galaxies {Top-Heavy IMFs in Starburst Galaxies}.
p.~125

\bibitem[\protect\citeauthoryear{{Scalo}}{{Scalo}}{1998}]{Scalo_98}
{Scalo} J.,  1998, in {Gilmore} G.,  {Howell} D.,  eds, ASP Conf. Ser. 142: The
  Stellar Initial Mass Function (38th Herstmonceux Conference) {The IMF
  Revisited: A Case for Variations}.
p.~201

\bibitem[\protect\citeauthoryear{Schartmann}{Schartmann}{2003}]{Schartmann_03}
Schartmann M., , 2003, Modelle f\"ur Staubtori in Aktiven Galaktischen Kernen

\bibitem[\protect\citeauthoryear{{Schartmann}}{{Schartmann}}{2007}]{Schartmann%
_07a}
{Schartmann} M.,  2007, PhD thesis, Max-Planck-Institute for Astronomy,
  Heidelberg, Germany

\bibitem[\protect\citeauthoryear{{Schartmann}, {Meisenheimer}, {Camenzind},
  {Wolf} \& {Henning}}{{Schartmann} et~al.}{2005}]{Schartmann_05}
{Schartmann} M.,  {Meisenheimer} K.,  {Camenzind} M.,  {Wolf} S.,    {Henning}
  T.,  2005, Astron. Astrophys., 437, 861

\bibitem[\protect\citeauthoryear{{Schartmann}, {Meisenheimer}, {Camenzind},
  {Wolf}, {Tristram} \& {Henning}}{{Schartmann} et~al.}{2008}]{Schartmann_08}
{Schartmann} M.,  {Meisenheimer} K.,  {Camenzind} M.,  {Wolf} S.,  {Tristram}
  K.~R.~W.,    {Henning} T.,  2008, Astron. Astrophys., 482, 67

\bibitem[\protect\citeauthoryear{{Shapiro} \& {Field}}{{Shapiro} \&
  {Field}}{1976}]{Shapiro_76}
{Shapiro} P.~R.,  {Field} G.~B.,  1976, Astrophys. J., 205, 762

\bibitem[\protect\citeauthoryear{{Shi}, {Rieke}, {Hines}, {Gorjian}, {Werner},
  {Cleary}, {Low}, {Smith} \& {Bouwman}}{{Shi} et~al.}{2006}]{Shi_06}
{Shi} Y.,  {Rieke} G.~H.,  {Hines} D.~C.,  {Gorjian} V.,  {Werner} M.~W.,
  {Cleary} K.,  {Low} F.~J.,  {Smith} P.~S.,    {Bouwman} J.,  2006, Astrophys.
  J., 653, 127

\bibitem[\protect\citeauthoryear{{Sodroski}, {Bennett}, {Boggess}, {Dwek},
  {Franz}, {Hauser}, {Kelsall}, {Moseley}, {Odegard}, {Silverberg} \&
  {Weiland}}{{Sodroski} et~al.}{1994}]{Sodroski_94}
{Sodroski} T.~J.,  {Bennett} C.,  {Boggess} N.,  {Dwek} E.,  {Franz} B.~A.,
  {Hauser} M.~G.,  {Kelsall} T.,  {Moseley} S.~H.,  {Odegard} N.,  {Silverberg}
  R.~F.,    {Weiland} J.~L.,  1994, Astrophys. J., 428, 638

\bibitem[\protect\citeauthoryear{{Stolte}, {Brandner}, {Grebel}, {Lenzen} \&
  {Lagrange}}{{Stolte} et~al.}{2005}]{Stolte_05}
{Stolte} A.,  {Brandner} W.,  {Grebel} E.~K.,  {Lenzen} R.,    {Lagrange}
  A.-M.,  2005, Astrophys. J., Lett., 628, L113

\bibitem[\protect\citeauthoryear{{Stolte}, {Grebel}, {Brandner} \&
  {Figer}}{{Stolte} et~al.}{2002}]{Stolte_02}
{Stolte} A.,  {Grebel} E.~K.,  {Brandner} W.,    {Figer} D.~F.,  2002, Astron.
  Astrophys., 394, 459

\bibitem[\protect\citeauthoryear{{Stone}, {Mihalas} \& {Norman}}{{Stone}
  et~al.}{1992}]{Stone_92c}
{Stone} J.~M.,  {Mihalas} D.,    {Norman} M.~L.,  1992, Astrophys. J., Suppl.
  Ser., 80, 819

\bibitem[\protect\citeauthoryear{{Stone} \& {Norman}}{{Stone} \&
  {Norman}}{1992a}]{Stone_92a}
{Stone} J.~M.,  {Norman} M.~L.,  1992a, Astrophys. J., Suppl. Ser., 80, 753

\bibitem[\protect\citeauthoryear{{Stone} \& {Norman}}{{Stone} \&
  {Norman}}{1992b}]{Stone_92b}
{Stone} J.~M.,  {Norman} M.~L.,  1992b, Astrophys. J., Suppl. Ser., 80, 791

\bibitem[\protect\citeauthoryear{{Sullivan}, {Le Borgne}, {Pritchet},
  {Hodsman}, {Neill}, {Howell}, {Carlberg}, {Astier}, {Aubourg}, {Balam},
  {Basa}, {Conley}, {Fabbro}, {Fouchez}, {Guy}, {Hook}, {Pain},
  {Palanque-Delabrouille}, {Perrett}, {Regnault}, {Rich}, {Taillet},
  {Baumont}, {Bronder}, {Ellis}, {Filiol}, {Lusset}, {Perlmutter}, 
  {Ripoche}, {Tao}}{Sullivan et~al.}{2006}]{Sullivan_06}
{Sullivan} M.,  {Le Borgne} D.,  {Pritchet} C.~J.,  {Hodsman} A.,  {Neill}
  J.~D.,  {Howell} D.~A.,  {Carlberg} R.~G.,  {Astier} P.,  {Aubourg} E.,
  {Balam} D.,  {Basa} S.,  {Conley} A.,  {Fabbro} S.,  {Fouchez} D.,  {Guy} J.,
   {Hook} I.,  {Pain} R.,  {Palanque-Delabrouille} N.,  {Perrett} K.,
  {Regnault} N.,  {Rich} J.,  {Taillet} R.,  {Baumont} S.,  {Bronder} J.,
  {Ellis} R.~S.,  {Filiol} M.,  {Lusset} V.,  {Perlmutter} S.,  {Ripoche} P.,
   {Tao} C.,  2006, Astrophys. J., 648, 868

\bibitem[\protect\citeauthoryear{{Sutherland} \& {Dopita}}{{Sutherland} \&
  {Dopita}}{1993}]{Sutherland_93}
{Sutherland} R.~S.,  {Dopita} M.~A.,  1993, Astrophys. J., Suppl. Ser., 88, 253

\bibitem[\protect\citeauthoryear{{Thompson}, {Quataert} \& {Murray}}{{Thompson}
  et~al.}{2005}]{Thompson_05}
{Thompson} T.~A.,  {Quataert} E.,    {Murray} N.,  2005, Astrophys. J., 630,
  167

\bibitem[\protect\citeauthoryear{{Tristram}, {Meisenheimer}, {Jaffe},
  {Schartmann}, {Rix}, {Leinert}, {Morel}, {Wittkowski}, {R{\"o}ttgering},
  {Perrin}, {Lopez}, {Raban}, {Cotton}, {Graser}, {Paresce} \&
  {Henning}}{{Tristram} et~al.}{2007}]{Tristram_07}
{Tristram} K.~R.~W.,  {Meisenheimer} K.,  {Jaffe} W.,  {Schartmann} M.,  {Rix}
  H.-W.,  {Leinert} C.,  {Morel} S.,  {Wittkowski} M.,  {R{\"o}ttgering} H.,
  {Perrin} G.,  {Lopez} B.,  {Raban} D.,  {Cotton} W.~D.,  {Graser} U.,
  {Paresce} F.,    {Henning} T.,  2007, Astron. Astrophys., 474, 837

\bibitem[\protect\citeauthoryear{{Urry} \& {Padovani}}{{Urry} \&
  {Padovani}}{1995}]{Urry_95}
{Urry} C.~M.,  {Padovani} P.,  1995, Publ. Astron. Soc. Pac., 107, 803

\bibitem[\protect\citeauthoryear{{van Leer}}{{van Leer}}{1977}]{Leer_77}
{van Leer} B.,  1977, Journal of Computational Physics, 23, 276

\bibitem[\protect\citeauthoryear{{V{\'a}zquez} \& {Leitherer}}{{V{\'a}zquez} \&
  {Leitherer}}{2005}]{Vazquez_05}
{V{\'a}zquez} G.~A.,  {Leitherer} C.,  2005, Astrophys. J., 621, 695

\bibitem[\protect\citeauthoryear{{Wada} \& {Norman}}{{Wada} \&
  {Norman}}{2002}]{Wada_02}
{Wada} K.,  {Norman} C.~A.,  2002, Astrophys. J., Lett., 566, L21

\bibitem[\protect\citeauthoryear{{Weedman}, {Hao}, {Higdon}, {Devost}, {Wu},
  {Charmandaris}, {Brandl}, {Bass} \& {Houck}}{{Weedman}
  et~al.}{2005}]{Weedman_05}
{Weedman} D.~W.,  {Hao} L.,  {Higdon} S.~J.~U.,  {Devost} D.,  {Wu} Y.,
  {Charmandaris} V.,  {Brandl} B.,  {Bass} E.,    {Houck} J.~R.,  2005,
  Astrophys. J., 633, 706

\bibitem[\protect\citeauthoryear{{Weidemann}}{{Weidemann}}{2000}]{Weidemann_00}
{Weidemann} V.,  2000, Astron. Astrophys., 363, 647

\bibitem[\protect\citeauthoryear{{Weingartner} \& {Draine}}{{Weingartner} \&
  {Draine}}{2001}]{Weingartner_01}
{Weingartner} J.~C.,  {Draine} B.~T.,  2001, Astrophys. J., 548, 296

\bibitem[\protect\citeauthoryear{{Wolf}}{{Wolf}}{2003}]{Wolf_03}
{Wolf} S.,  2003, Computer Physics Communications, 150, 99

\bibitem[\protect\citeauthoryear{{Wolf}, {Henning} \& {Stecklum}}{{Wolf}
  et~al.}{1999}]{Wolf_99a}
{Wolf} S.,  {Henning} T.,    {Stecklum} B.,  1999, Astron. Astrophys., 349, 839

\end{thebibliography}

\bsp

\label{lastpage}

\end{document}